\documentclass[a4paper,11pt]{article}
\usepackage{jinstpub} 
\usepackage{lineno}
\usepackage{algorithm}
\usepackage{subcaption}
\usepackage[noend]{algpseudocode}
\usepackage{xcolor}
\newcommand{\corrauth}{\textcolor{blue}{1}}

\newif\ifwithlineno

\title{ Introducing a Markov Chain-Based Time Calibration Procedure for Multi-Channel Particle Detectors: Application to the SuperFGD and ToF Detectors of the T2K Experiment }

\author[a]{S.~Abe}
\author[b]{H.~Alarakia-Charles}
\author[c]{I.~Alekseev}
\author[s]{C.~Alt}
\author[d]{T.~Arai}
\author[e]{T.~Arihara}
\author[f]{S.~Arimoto}
\author[g]{A.M.~Artikov}
\author[e]{Y.~Awataguchi}
\author[h]{N.~Babu}
% ...
\ifwithlineno
\else
\author[ ]{\vspace*{\fill}\newpage}
\fi

\affiliation[a]{University of Tokyo, Institute for Cosmic Ray Research, Kamioka Observatory, Kamioka, Japan}
\affiliation[b]{Lancaster University, Physics Department, Lancaster, United Kingdom}
\affiliation[c]{Lebedev Physical Institute of the Russian Academy of Sciences, Moscow, Russia}
\affiliation[d]{University of Tokyo, Tokyo, Japan}
\affiliation[e]{Tokyo Metropolitan University, Department of Physics, Tokyo, Japan}
\affiliation[f]{Kyoto University, Department of Physics, Kyoto, Japan}
\affiliation[g]{Joint Institute for Nuclear Research, Dubna, Moscow Region, Russia}
\affiliation[h]{Louisiana State University, Baton Rouge, USA}
% ...
\affiliation[s]{Institute for Particle Physics and Astrophysics, ETH Zurich, Zurich, Switzerland}
% ...

\emailAdd{lorenzo.giannessi@unige.ch}
\emailAdd{vedantha.kasturi@unige.ch}
\emailAdd{greina@uni-mainz.de}
\emailAdd{haowei.zheng@stonybrook.edu}
\note[\corrauth]{Corresponding author(s).}

\abstract{
Inter-channel mis-synchronisation can be a limiting factor to the time resolution of high performance timing detectors with multiple readout channels and independent electronics units. In these systems, time calibration methods employed must be able to efficiently correct for minimal mis-synchronisation between channels and achieve the best detector performance.
We present an iterative time calibration method based on Markov Chains, suitable for detector systems with multiple readout channels. Starting from correlated hit pairs alone, and without requiring an external reference time measurement, the method solves for fixed per-channel offsets, with precision limited only by the intrinsic single-channel resolution. A mathematical proof that the method is able to find the correct time offsets to be assigned to each detector channel in order to achieve inter-channel synchronisation is given, and it is shown that the number of iterations to reach convergence within the desired precision is controllable with a single parameter. Numerical studies are used to confirm unbiased recovery of true offsets. Finally, the application of the calibration method to the Super Fine-Grained Detector (SuperFGD) and the Time of Flight (TOF) detector at the upgraded T2K near detector (ND280) shows good improvement in overall timing resolution, demonstrating the effectiveness in a real-world scenario and scalability.
}


\arxivnumber{2508.07846} 

\begin{document}

\maketitle

\section{Introduction}
Large-scale ionisation detectors typically employ a substantial number of readout channels organised into multiple independent electronics units, synchronised through the use of a common primary clock signal. Contemporary examples can be found in scintillator detectors used as active targets for neutrino experiments, where light is conveyed to the optical sensor via optical fibres~\cite{T2KND280FGD:2012umz,Yamamoto:2004pvi,MINERvA:2006aa,MINOS:1998kez,NOvA:2007rmc,BabyMIND:2017mys}. In these systems, secondary clock phase-locking mis-synchronisation, imperfect cable length and trace matching, and comparator switching delays can result in fixed offsets in the hit times recorded by different channels that can negatively affect the resultant time resolution of the detector. Great progress is being made to reduce the time resolution of individual channels, both in terms of enhancing the readout electronics performance, and in the development of fast scintillating materials~\cite{Andr__2025,FERRI2020162159,Cavallari_2020}. As the time resolution of individual readout channels improves, it becomes imperative to address the mis-synchronisation between multiple channels, in order to prevent that from becoming the limiting factor in a detector's resultant time resolution.
Therefore, it is common practice in these detector systems to perform a time calibration in order to compensate for inter-channel mis-synchronisation.
This consists of finding a fixed time offset $T^{(0)}$, for each readout channel, such that:
\begin{equation}
    t_{\text{meas}} + T^{(0)} = t_{\text{exp}} = t_{\text{ref}} + t_{\text{transit}},
    \label{eq:def_time_offset}
\end{equation}
where $t_{\text{meas}}$ is the time measured by the channel, $t_{\text{exp}}$ is  the expected time that the signal was created, $t_{\text{transit}}$ is the time expected to elapse between the reference time $t_{\text{ref}}$ and the recording of the signal, depending on the detector geometry, light propagation, and light readout methods. This traditional method requires the estimation of the reference time through the reconstruction of ionising particle tracks in the detector using uncalibrated data~\cite{antares}. This adds complexity to the problem and potentially introduces track reconstruction biases, which can be avoided with the method we describe in section~\ref{sec:iterative compensation}, due to its iterative nature and independence from a fixed reference time. 

We present a general calibration method that exploits the correlation between different channels of the detector, and iteratively finds the time offsets for all the detector channels without requiring any reference time, making the time offset calibration computationally easier and unbiased.

Similar concepts of exploiting pairs of hits in different detector channels have been used in other contexts as a time calibration method~\cite{PET1,PET2}; in this work, we present the mathematical principle on which our time calibration method stands, demonstrating its consistency and verifying its effectiveness by means of numerical simulations. We then present two real-life applications in high energy physics experiments where this method is used to enhance the time resolution performance. The applications presented are in the context of the T2K Near Detector (ND280):
a fine-grained scintillating cubes based active neutrino target (Super Fine-Grained Detector, or SuperFGD), and a time-of-flight scintillating bars detector (ToF)~\cite{Blondel:2017orl,Mineev:2018ekk,T2K:2019bbb}. 
\label{sec:intro}
\section{Pairs of hits highly correlated in time}
The method presented in this manuscript is based on the signal correlation between different readout channels in the detector. It is therefore important to introduce the concept of highly correlated hit pairs, or \textit{`matching hit pairs'}. When an ionising particle crosses a detector, it typically generates a signal (a `hit') in multiple readout channels. Normally, if the same particle leaves a signal in two channels $ch_1$ and $ch_2$, respectively, at times $t_1$ and $t_2$, it is possible to compute the expected time difference between the two hits based on simple geometrical arguments, and without reconstructing the track inside the detector. This can be done, for example, by knowing the position of the readout units, or by exploiting the adjacency of scintillating units read by different channels. If an estimation of the expected time difference can be performed for a pair of hits, the times of the two hits are highly correlated, and the pair is designated as a \textit{matching hit pair}.

In the ideal case where each single detector time measurement is unbiased, there is no discrepancy between the measured and expected time difference between matching hits. The calibration method here proposed is based on the global minimisation of the discrepancy between measured and expected time differences over a sample of hit pairs measured by any pair of readout channels in the detector. We define this discrepancy as $\Delta$: 
\begin{equation}
    \Delta = \Delta t_{\text{meas}}-\Delta t_{\text{exp}} = (t_{1}-t_{2}) - (s_{1}-s_{2}),
    \label{eq:delta}
\end{equation}
where $t_{1}$ and $t_{2}$ are the measured hit times, and $s_{1}$ and $s_{2}$ are the expected hit times, for the first and second hits, respectively, from a \textit{matching hit pair}. The global minimisation of $\Delta$ is performed iteratively and simultaneously for all channels, as outlined in the following.

It is important to stress that, throughout the treatment presented here, the measured times and the expected times always appear respectively in the differences $t_{1}-t_{2}$ and $s_{1}-s_{2}$, eliminating the need for a global reference time estimation.

\subsection{Matching hits in the SuperFGD}
The SuperFGD is an active neutrino target composed of nearly two million optically isolated scintillating cubes of 1~cm side. It has 3D tracking and calorimetry capabilities and was recently installed in the near detector of the T2K experiment~\cite{Blondel:2017orl}. The scintillating material is polystyrene doped with 1.5~\% paraterphenyl (PTP) and 0.01~\% POPOP. 

\begin{figure}
    \centering
    \includegraphics[width=0.5\linewidth]{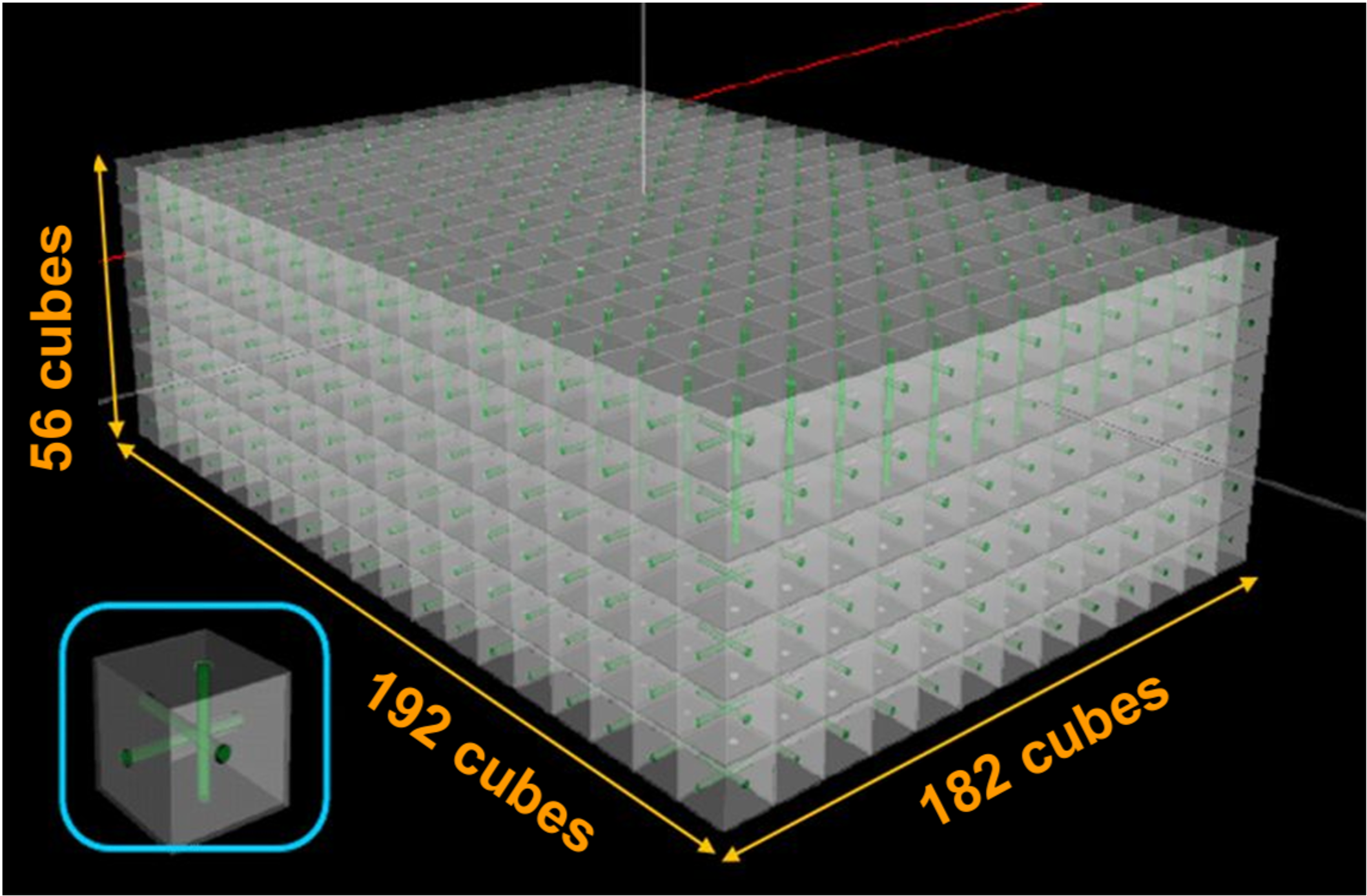}
    \caption{A schematic 3D rendering of the SuperFGD detector. Light readout happens on one end of each fibre, on the surface of the detector.}
    \label{fig:SFGD}
\end{figure}

Each cube is crossed by three wavelength-shifting (WLS) optical fibres, oriented along the three Cartesian directions, that guide the light to the surface of the detector, where Silicon PhotoMultipliers (SiPMs) read the light signals out. The fibres cross the entire detector so that each fibre can carry the light generated in multiple cubes. A scheme of the detector geometry is provided in Figure~\ref{fig:SFGD}. The importance of having an accurate time resolution in the SuperFGD is primarily motivated by the physics goal of measuring the kinetic energy of neutrons produced in neutrino-nucleus interactions via time-of-flight measurements within the detector volume.

\begin{figure}
    \centering
    \includegraphics[width=0.33\linewidth]{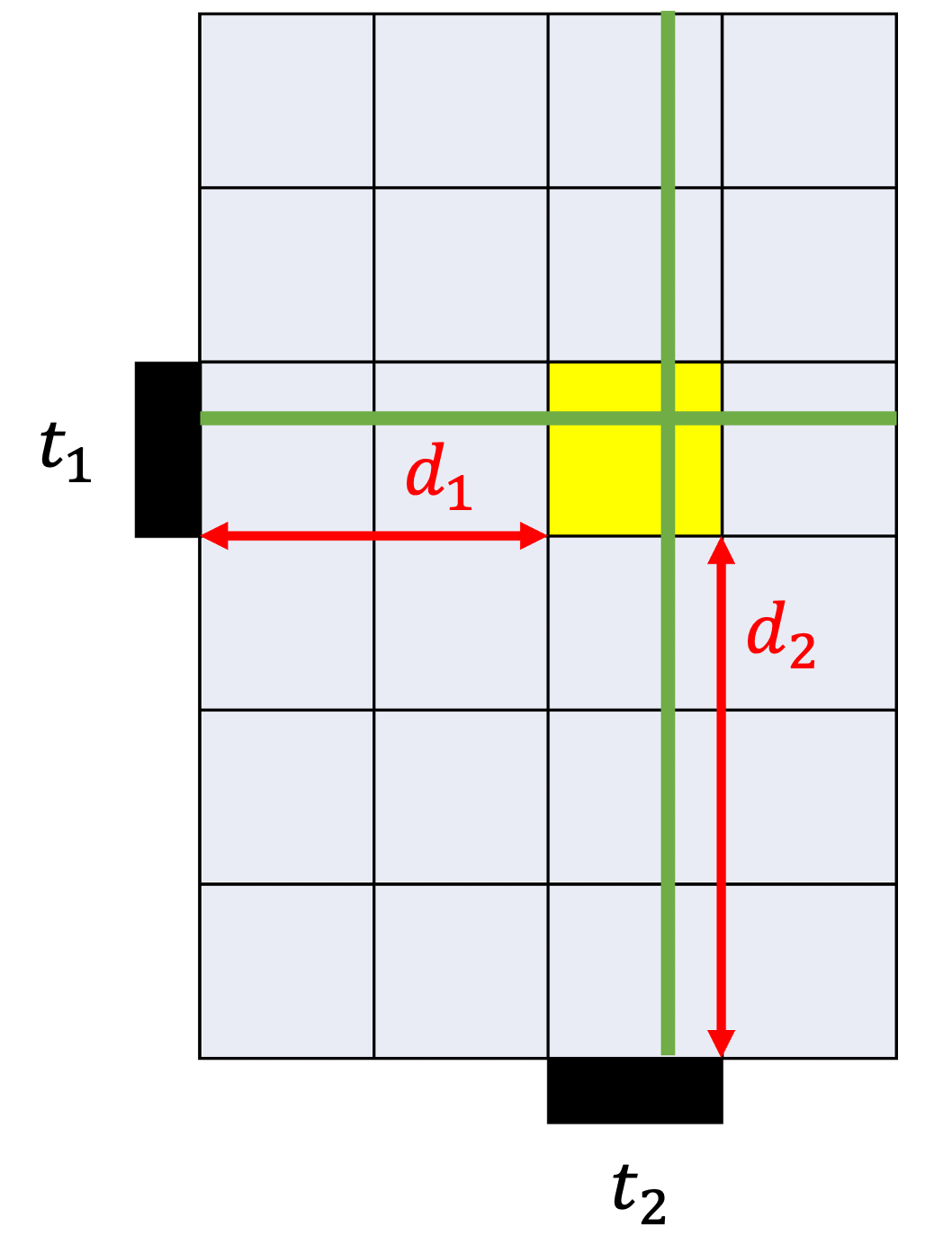}
    \caption{Schematic 2D view of a matching hit pair in a SuperFGD-like setup. The scintillating light is generated in the yellow cube and propagates to the SiPM (black boxes) along the optical fibres (green). The times $t_1$ and $t_2$ are measured at the arrival of the scintillating light on the SiPMs, and the distances $d_1$ and $d_2$ are estimated by geometry.} 
    \label{fig:sfgd_hit_pair}
\end{figure}

The geometry of the SuperFGD offers a unique opportunity to observe pairs of hits that are highly correlated in time: when scintillating light is generated in an individual cube, it may be captured by the three optical fibres and travel along them to the light sensors, generating a hit. 
Thanks to the intersection of the optical fibres, it is possible to precisely estimate the location of the cube where the light was produced. Therefore, the distance from the scintillation light emission point to the SiPM is known, allowing us to calculate the expected time differences between the three pairs of hits. This concept is illustrated schematically in Figure~\ref{fig:sfgd_hit_pair}.

The hit times $t_1$ and $t_2$ are independently measured, and the expected time difference is known very accurately, with a precision comparable to with the width of a cube over the speed of light in the fibre. The quantity $\Delta$ can be easily estimated for each hit pair as: 
\begin{equation}
    \Delta = (t_1-t_2)-\dfrac{(d_1-d_2)}{v_\text{{fibre}}},
    \label{eq:sfgd_hit_pair}
\end{equation}
where $d_1$ and $d_2$ are the distances from the cube to the two SiPMs, and $v_\text{{fibre}}$ is the propagation speed of light in the optical fibre.

It must be noted that in a more realistic scenario, the scintillating light is sometimes generated in more than one single cube along one fibre. Therefore, the distance from the generation point of the light to the SiPMs can be estimated with a precision of a few cube widths, resulting in a slight smearing in the expected time difference estimation. This effect can be minimised by selecting the dataset of matching hit pairs used for calibration in such a way that whenever a matching hit is found in a certain channel, the number of cubes lighting up along the corresponding optical fibre is not too large (depending on the desired precision).
It must be noted that according to this definition, the speed of propagation of light in the fibre is assumed to be constant. More generally, optical effects can change the effective velocity of light propagation in the fibre. Deviations of the effective propagation velocity from its average can be measured as a function of the distance to the SiPM, and more accurate results can be obtained by modifying equation~\ref{eq:sfgd_hit_pair}:
\begin{equation}
    \Delta = (t_1-t_2)-\left(\int_{d_1}^0\dfrac{dx}{v_\text{{fibre}}(x)}-\int_{d_2}^0\dfrac{dx}{v_\text{{fibre}}(x)}\right),
    \label{eq:sfgd_hit_pair_integral}
\end{equation}
where $v_\text{{fibre}}(x)$ is the speed of light in the fibre as a function of the distance from the SiPM.

\subsection{Matching hits in the ToF detector} 
\begin{figure}[h]
    \centering
    \includegraphics[width=0.5\linewidth]{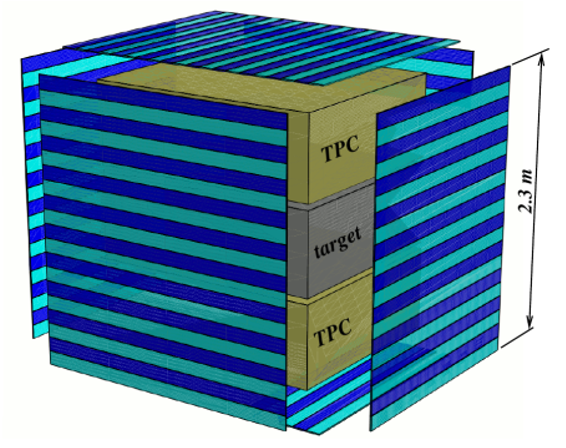}
    \caption{Exploded view of the upstream part of the ND280 detector. The target is the SuperFGD. The ToF detector scintillating bars enclose the SuperFGD and two Time Projection Chambers.}
    \label{fig:tof-scheme}
\end{figure}

The ToF detector~\cite{Korzenev:2021mny} was also installed as part of the T2K near detector upgrade and consists of 118 2~m-long scintillator bars composed of EJ-200 material, arranged into six panels which enclose the recently upgraded ND280 upstream tracking detectors (Figure~\ref{fig:tof-scheme}). These provide precise time measurements for tracks entering or exiting the tracking detectors and can veto out-of-fiducial background interactions.

\begin{figure}
    \centering
    \includegraphics[width=0.5\linewidth]{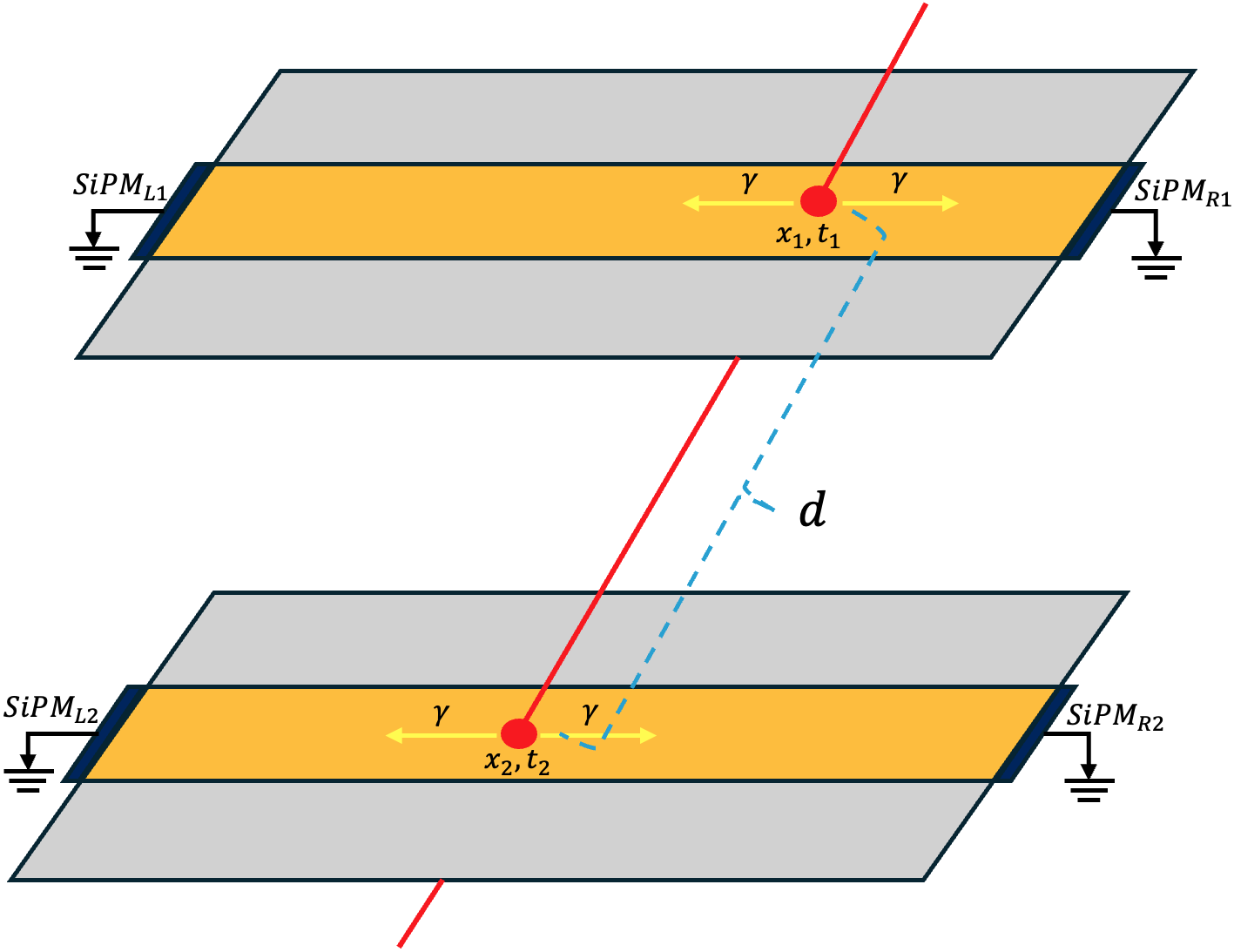}
    \caption{Schematic illustration of a pair of highly correlated hit pairs in the ToF detector. We consider here an arbitrary pair of ToF bars. The times and positions are measured using the dual readout of scintillating light on the two ends of each bar.}
    \label{fig:tof_hit_pair}
\end{figure}

The scintillating light generated in the bars is read by SiPMs on both ends of each bar, allowing time and position measurements for hits generated by ionising particles. In this configuration, when a particle crosses two scintillating bars located in two different panels, the expected time difference between the two hits generated can be easily estimated knowing the hit positions, allowing the $\Delta$ to be computed as:  

\begin{equation}
    \Delta = t_1 - t_2 - \dfrac{d}{c},
    \label{Tof_delta_time_eq}
\end{equation}
where, $ t_i = \dfrac{t_L + t_R}{2} - \dfrac{L}{2v}$ with $i = 1, 2$  are the times measured by each of the bars in the separate panels, the $t_L , t_R $ are the actual times of the hits in the SiPM on each end of an individual bar, $v$ the speed of light propagation within the scintillator bar,  $c$ is the speed of the ionising particle, approximated with the speed of light in vacuum, $d$ the distance between the hit positions, and $L$ is the length of the bar. The distance is computed by the reconstruction of the entry point $x_i$ of the ionising particle along the bar based on the averaged light propagation times from the reconstructed particle crossing times $t_i$. This is illustrated in Figure~\ref{fig:tof_hit_pair}.
There are two ways of reconstructing the position $x_i$ of interaction along the bar in the ToF detector. One method is performed by the ToF detector alone, and it relies on the dual-ended time measurement on the two ends of the bar ($t_L, t_R$). In this case, it is necessary to know the effective speed of propagation of light inside the scintillating bar. Alternatively, an external tracking detector (such as the SuperFGD or the High-Angle TPCs in the context of ND280) can be used to provide a precise position reference.

\subsection{Matching hits in scintillating bar detectors}

Another common detector geometry where the concept of matching hit pairs can be applied is that of scintillating bar detectors with multiple planes of scintillating bars along different directions, similar to the detector geometry of MINER$\nu$A~\cite{MINERvA:2006aa} or the T2K FGDs~\cite{AMAUDRUZ20121}.

\begin{figure}[!h]
\centering
    \centering
    \includegraphics[width=0.4\linewidth]{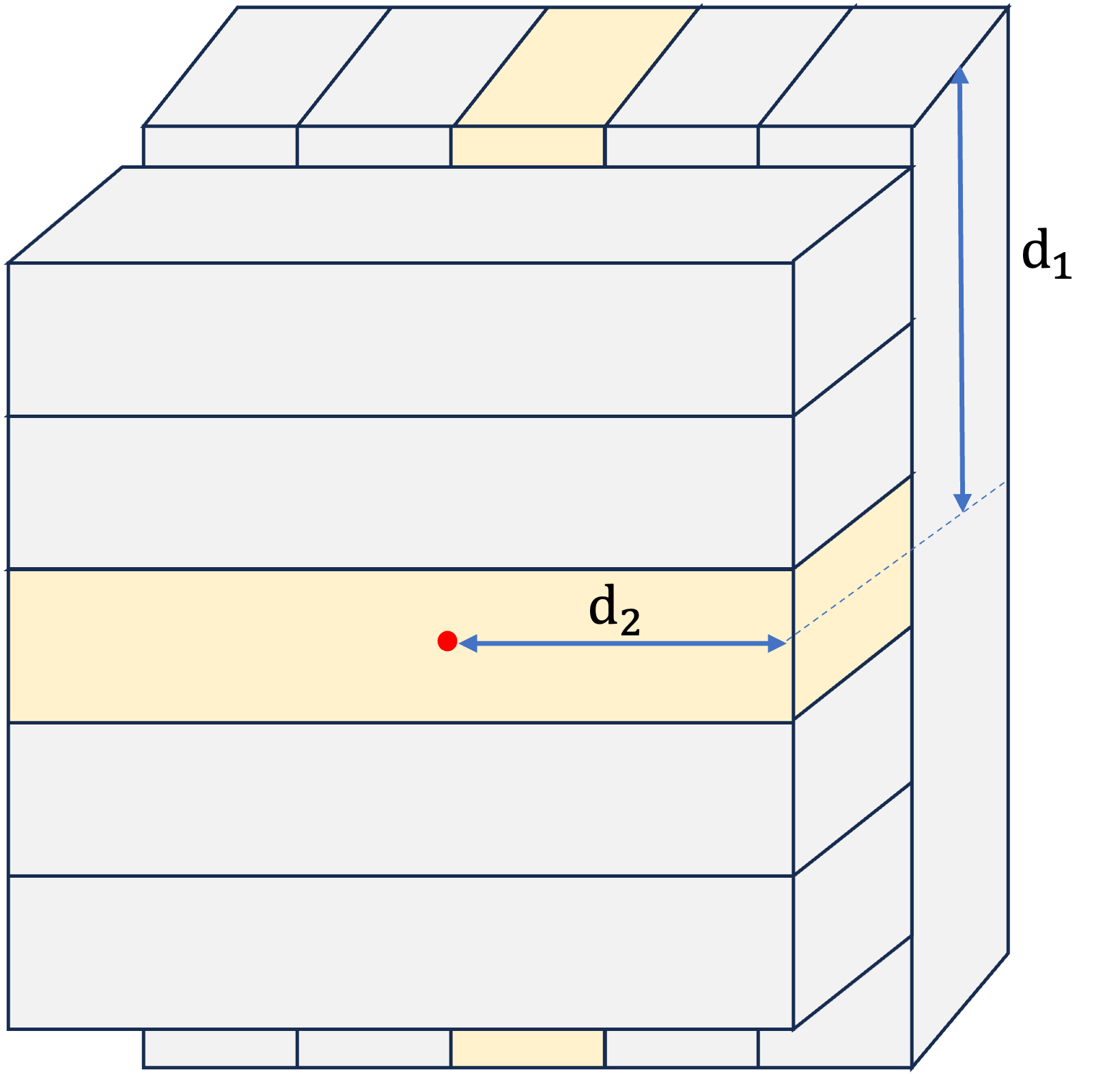}
    \caption{A matching hit pair in adjacent scintillating planes in a generic ``2D'' scintillating bar detector. Here, $d_1$ and $d_2$ represent the distances from the light sensor (typically installed on one end of the bar) to the interaction point estimated as the point where the two bars with detected signals cross.}
    \label{fig:fgd_hit_pair}
\end{figure}

Pairs of matching hits can be built in these detectors from hits generated in two adjacent planes, in bars aligned along different directions. The position along the bars can be estimated thanks to the conjugate hit in the adjacent plane (Figure~\ref{fig:fgd_hit_pair}), locating the scintillation light emission point with a precision of the order of the bar thickness. In this case, the definition of $\Delta$ resembles that of the SuperFGD:
   
\begin{equation}
    \Delta = t_1 - t_2 - \dfrac{d_1-d_2}{v_\text{{fibre}}}.
    \label{minerva_delta_time_eq}
\end{equation}

\label{sec:matching_hits}

\section{Iterative time offset compensation}
\label{sec:iterative compensation}
As introduced in section~\ref{sec:intro}, the problem of calibrating the time offsets of a detector can be presented as the problem of finding a vector $\mathbf{T}^{(0)}$ of dimension $N_{ch}$ equal to the number of channels in the detector: 
\begin{align}
    \mathbf{T}^{(0)} &= \begin{pmatrix}
           T^{(0)}_{1} \\
           T^{(0)}_{2} \\
           \vdots \\
           T^{(0)}_{N_{ch}}
         \end{pmatrix}
    \label{eq:def_deltat0_vector}
\end{align}
that corrects the measured time of each channel.  
That is, given a certain hit in channel $i$, with measured time $t_i$ and expected time $s_i$: 
\begin{equation}
    T_{i}^{(0)} + t_i = s_i,
    \label{eq:single_hit_time_offset}
\end{equation}
where, according to our assumptions, $T_{i}^{(0)}$ is a function only of the readout channel. 
Given that the detector has an intrinsic non-zero time resolution, this statement is true only on average,  with the standard deviation of $\Delta$ expressing a measure of the detector time resolution after time calibration.

\subsection{Proof of algorithm convergence}
Let us now assume that a certain number $N$ of matching hit pairs has been selected for a given detector geometry.

In the following, we denote with $t_{1,n}$ and $t_{2,n}$ the first and second raw measured times of the $n$-th matching hit pair in the dataset, and with $t_{1,n}^{(k)}$ and $t_{2,n}^{(k)}$ the corrected hit times after $k$ iterations, while $s_{1,n}$ and $s_{2,n}$ are the respective expected hit times. Notice that the individual expected times $s_{1,n}$ and $s_{2,n}$ are unknown, but their difference can be estimated as outlined in section~\ref{sec:matching_hits}.

For the $n$-th hit pair we can compute $\Delta$ as:
\begin{equation}
    \Delta =  \Delta t_{n}-\Delta s_{n} = (t_{1,n}-t_{2,n})-(s_{1,n}-s_{2,n}).
    \label{eq:delta_n_definition}
\end{equation}

The time offset calibration method consists of an iterative correction of the measured times in each channel of the detector. The correction at iteration $k$ is computed using the times as corrected after $k-1$ iterations, with the times at iteration 0 being the raw measured times. Finally, the time offset for a certain channel is found as the sum of all the corrections computed at each iteration. We will demonstrate under general assumptions that this procedure (with the correct choice of correction) converges to a unique vector of time offsets.

We define the correction at iteration $k$ for a certain channel $i$ as the average of $\Delta_n$ over all matching hit pairs such that the channel of the first hit in the matching hit pair $n$ is $i$:
\begin{equation}
    \Delta^{(k)}_{i} = \dfrac{1}{N_{i}}\sum_{i}\Delta^{(k)} = \dfrac{1}{N_i}\sum_{i} \left[ \left( t^{(k)}_{1,n}-t^{(k)}_{2,n} \right) - \left( s_{1,n}-s_{2,n} \right) \right].
    \label{eq:correction}
\end{equation}
Moreover, we defined the symbol: 
\begin{equation*}
\sum_i := \sum_{\substack{n=1\\ch_1=i}}^{N_{hits}},
\end{equation*}
as the sum over all the matching hits such that the first channel in the hit $n$ is channel $i$.
In order not to give relevance to the place that a hit occupies in the matching hit pair, the dataset is duplicated: for each hit pair in the dataset, we also evaluate the hit pair where the first and second hits are swapped. This makes the problem symmetric with respect to the transformation $1 \leftrightarrow 2$ in any hit pair. The hit times are updated at each iteration as: 
\begin{equation}
    t^{(k+1)}_{i} = t^{(k)}_{i} -\alpha\Delta^{(k)}_{i},
    \label{eq:iteration}
\end{equation}
where $\alpha$ is a damping factor, defined arbitrarily between 0 and 1. Later, it will be shown that $\alpha$ determines the speed with which the algorithm converges. 
According to our assumption (equation~\ref{eq:single_hit_time_offset}), the only difference between the expected and measured time difference of a hit in a determined channel is due to a channel-dependent time offset. Therefore, in any given iteration $k$, the correction $\alpha\Delta^{k}_{i}$ depends only on the corresponding channel $i$, and the \textit{residual time offset} $T^{(k)} = t^{(k)} - s $ is characteristic only of the readout channel. With this assumption, we can define a vector $\mathbf{T}^{(k)}$ with $N_{ch}$ elements, equivalent to the $\mathbf{T}^{(0)}$ introduced in equations~\ref{eq:def_deltat0_vector} and~\ref{eq:single_hit_time_offset}, that is computed with the corrected times $t^{(k)}$ after $k$ iterations.

The vector of the residual time offsets $\mathbf{T}^{(k)}$ represents the average difference between the measured time, corrected after $k$ iterations, and the expected time of the hits recorded in a given channel. Therefore, performing the time calibration is equivalent to finding an iterative process such that: 
\begin{equation}
      \lim_{k\rightarrow\infty}\mathbf{T}^{(k)}=\begin{pmatrix}
    0 \\
    \vdots \\
    0 
    \end{pmatrix}.
    \label{eq:residual_offset_goes_to_0}      
\end{equation}
In practice, the time calibration is complete when the elements of the vector $\mathbf{T}^{(k)}$ are much smaller than the time resolution of the single readout channel for a certain iteration $\mathcal{K}$. If this is achieved, one can estimate the time offset of a given channel $T^{(0)}_{i}$ simply as the sum $W_i$ of all the corrections applied on $i$ at every iteration before reaching convergence:
\begin{equation}
    W_{i} := \sum_{k=0}^{\mathcal{K}} \alpha\Delta^{(k)}_{i},
    \label{eq:defOffsetEstimation}
\end{equation}
or more compactly for every channel:
\begin{equation}
    \mathbf{W} := \sum_{k=0}^{\mathcal{K}} \alpha\mathbf{\Delta}^{(k)}.
    \label{eq:defOffsetEstimation_vector}
\end{equation}
We are going to demonstrate that condition~\ref{eq:residual_offset_goes_to_0} holds, and $\mathbf{W}=\mathbf{T}^{(0)}$ in the limit of many iterations.

Let us consider the iterative corrections on a target channel $i$ to calibrate and compute $T^{(k)}_{i}$ as a function of the times computed at the previous iterations:
\begin{equation}
    T^{(k)}_{i}=t^{(k)}_{i}-s_{i}.
\end{equation}
By using the times corrected at the previous iteration we obtain:
\begin{align}
    T^{(k)}_{i} =& t^{(k)}_{i} - s_{i} \nonumber\\
    =& t^{(k-1)}_{i} - \alpha\Delta^{(k-1)}_{i} - s_{i} \nonumber \\
    =& T^{(k-1)}_{i} - \alpha\Delta^{(k-1)}_{i},
\label{eq:correction_on_deltaT}
\end{align}
where we omitted the subscript relative to the matching hit to lighten the notation.
By using equation~\ref{eq:correction}, we obtain:
\begin{equation}
    T^{(k)}_{i} = T^{(k-1)}_{i} - \alpha\left[ \dfrac{1}{N_i}\sum_i(t^{(k-1)}_{1,n}-s_{1,n}) - \dfrac{1}{N_i}\sum_i(t^{(k-1)}_{2,n}-s_{2,n})  \right].
    \label{eq:residual_offset}
\end{equation}
The first sum is over all matching hits where the first hit channel is channel $i$, which by definition means that all the terms in the first sum are equal to $T^{(k-1)}_{i}$. Therefore:
\begin{equation}
    T^{(k)}_{i} = T^{(k-1)}_{i} - \alpha\left[ T^{(k-1)}_{i}  - \dfrac{1}{N_i}\sum_i(t^{(k-1)}_{2,n}-s_{2,n})  \right].
\end{equation}

We then label $i\times$ any channel that can generate a matching hit pair with channel $i$ acting as the other component of the pair. We denote these channels as \emph{$i$-crossing}. It is evident that only the geometry of the detector defines this property of the readout channels.

By this definition, all the channels involved in the second sum of equation~\ref{eq:residual_offset} are ``$i$-crossing'' channels. We can finally write:
\begin{equation}
    T^{(k)}_{i} = T^{(k-1)}_{i} - \alpha T^{(k-1)}_{i} + \dfrac{\alpha}{N_i}\sum_i T^{(k-1)}_{i\times}.
    \label{eq:deltaT_i_final_scalar}
\end{equation}
Equation~\ref{eq:deltaT_i_final_scalar} shows that the correction to the hit time applied at each iteration contains information about the residual time offset of the target channel, as well as an additional term generated by the yet unknown residual offsets of the crossing channels. 
We can substantially simplify equation~\ref{eq:deltaT_i_final_scalar} by noticing that it can be written in matrix form:

\begin{equation}
    T^{(k)}_{i} = M_{ij} T^{(k-1)}_{j}.
\end{equation}
It is now necessary to define the elements of the matrix $M$ explicitly. To this end, we introduce another more fundamental matrix, the \textit{crossing matrix} $\mathbb{X}$.

\subsection{The crossing matrix} To build the crossing matrix we start with a $N_{ch}\times N_{ch}$ symmetric matrix where the element $ij$ is 1 if channel $i$ can form a matching hit pair with channel $j$ (channels $i$ and $j$ are ``\textit{crossing channels}''). This matrix is characteristic only of the way in which a matching hit pair is defined (the detector geometry), and maps the vector of residual offsets $\mathbf{T}^{(k)}$ to the sum of residual offsets on the crossing channels. An additional evident property is that the diagonal elements are null.

In a general matching hit pair dataset, there is an arbitrary number $w_{ij}$ of matching hit pairs between channel $i$ and $j$. Hence, the elements of the crossing matrix $\mathbb{X}$ defining a specific matching hit pair dataset are such that 
\begin{equation}
    x_{ij}= 
\begin{cases}
    \dfrac{w_{ij}}{\sum_j w_{ij}}   &   \text{if } \text{channel }i\text{ crosses channel }j\\
    0             & \text{otherwise}
\end{cases},
    \label{eq:crossing_matrix_def}
\end{equation}
where $w_{ij}$ is a weight that describes the probability of finding a matching hit pair with channels $(i,j)$ in our dataset. In this case, $\mathbb{X}$ is no longer symmetric, and is characteristic of the specific matching hit pairs dataset. $\mathbb{X}$ will depend on the conditions in which the dataset was acquired and the hit pairs selected. In a good calibration dataset (i.e.\ one where the detector is uniformly hit), the weights $w_{ij}$ are all similar to each other, and the matrix depends on the dataset only marginally, retaining the information of the detector geometry. In order to efficiently calibrate all channels, it is important to accurately select the dataset in such a way that no particular region or set of channels are more likely to be hit. If this is the case:
\begin{equation}
    \dfrac{w_{ij}}{\sum_j w_{ij}} \simeq \dfrac{1}{N_i},
\end{equation}
where $N_i$ is the number of channels crossing channel $i$, a purely geometrical property. Using the definition~\ref{eq:crossing_matrix_def}, equation~\ref{eq:deltaT_i_final_scalar} can be written as
\begin{equation}
    \mathbf{T}^{(k)} = M \mathbf{T}^{(k-1)} = [(1-\alpha)\mathbb{I} +\alpha \mathbb{X}]\mathbf{T}^{(k-1)}
    \label{eq:deltat_i_function_of_deltat_iminus1}.
\end{equation}
Therefore:
\begin{equation}
    \mathbf{T}^{(k)} = [(1-\alpha)\mathbb{I} +\alpha \mathbb{X}]^k\mathbf{T}^{(0)}.
    \label{eq:deltat_i_function_of_deltat_0}
\end{equation}
By explicitly writing down the elements of matrix $M$ in equation~\ref{eq:deltat_i_function_of_deltat_iminus1},
\begin{equation}
    m_{ij}=\\
\begin{cases}
    1-\alpha    &   \text{if } i=j\\
    \alpha\dfrac{w_{ij}}{\sum_j w_{ij}}   &   \text{if } \text{channel }i\text{ crosses channel }j\\
    0             & \text{otherwise}
    \label{eq:definition_M_matrix}
\end{cases}
\end{equation}
it is observed that all the elements of $M$ are positive, and 
$$\sum_j m_{ij}=1\hspace{5mm}\forall i.$$
It follows that $M$ is a row-stochastic Markov matrix, and it allows us to find an analogy between the iterative time calibration procedure and a Markov chain~\cite{Markov} process with transition matrix $M$.

From one of the fundamental theorems describing the algebra of Markov chains, by Perron and Frobenius~\cite{Perron-Frobenius}, under mostly general conditions (see appendix~\ref{sec:irreducible} for more details) the eigenvalues of a row-stochastic matrix $\lambda_{i}$ respect the following condition:
\begin{equation}
    1 = \lambda_{0} > \left|\lambda_{1}\right| \ge \left|\lambda_{2}\right| \ge ... \ge \left|\lambda_{n}\right|.
    \label{eq:eigenvalues}
\end{equation}
Moreover, the eigenvector $\mathbf{e}_{0}$ corresponding to the largest eigenvalue $\lambda_{0}$=1 is, by construction, the 1-s vector $\mathbf{1}_{N_{ch}}$ of size $N_{ch}$, such that
\begin{equation}
    M\mathbf{e}_{0} = \lambda_{0}\mathbf{e}_{0}=\lambda_{0}\begin{pmatrix}
    1 \\
    \vdots \\
    1 
    \end{pmatrix} .
\label{eq:e0}
\end{equation}
From equation~\ref{eq:deltat_i_function_of_deltat_0} we have:
\begin{equation}
    \mathbf{T}^{(k)} = M^k\mathbf{T}^{(0)}.\\
    \label{eq:matrix-form}
\end{equation}
According to the \textit{power iteration} algorithm\footnote{Sometimes called simply ``power method''}~\cite{powermethod}, for any non-zero vector $\mathbf{T}^{(0)}$, in the limit of a large number of iterations, the sequence in equation~\ref{eq:matrix-form} converges to the eigenvector of $M$ associated to its largest eigenvalue, hence $\mathbf{1}_{N_{ch}}$.

This can be simply shown by expressing $\mathbf{T}^{(0)}$ as a linear combination of $\mathbf{e}_{i}$, basis of $\mathbb{C}^{N_{ch}}$ obtained from the generalised eigenvectors of $M$. This is possible in $\mathbb{C}^{N_{ch}}$ even if $M$ is not diagonalisable~\cite{saad}, and one vector of the basis is guaranteed to be $\mathbf{e}_{0}=\mathbf{1}_{N_{ch}}$. We can then express
\begin{equation}
    \mathbf{T}^{(0)} = \sum_{i=0}^{N_{ch}}c_{i}\mathbf{e}_{i},
    \label{eq:T0-linear-combination}
\end{equation}
where, $c_{i}$ are scalar coefficients $\in \mathbb{C}$.

Therefore, equation~\ref{eq:matrix-form} can be rewritten as:
\begin{equation}
\begin{split}
    \mathbf{T}^{(k)}=M^{k}\sum_{i=0}^{N_{ch}}c_{i}\mathbf{e}_{i} = \sum_{i=0}^{N_{ch}}c_{i}M^{k}\mathbf{e}_{i} =\sum_{i=0}^{N_{ch}}c_{i}\lambda_{i}^{k}\mathbf{e}_{i} = \\\lambda_{0}^{k}c_{0}\mathbf{e}_{0} + \sum_{i=1}^{N_{ch}}c_{i}\lambda_{i}^{k}\mathbf{e}_{i}= \lambda_{0}^{k}\left[c_{0}\mathbf{e}_{0} + \sum_{i=1}^{N_{ch}}c_{i}\left(\frac{\lambda_{i}}{\lambda_{0}}\right)^{k}\mathbf{e}_{i}\right]      .  
\end{split}
\label{eq:Tklimit}
\end{equation}
This, in the limit of $k\rightarrow\infty$, tends to $c_{0}\mathbf{e}_{0}$.
We conclude that
\begin{equation}
    \lim_{k\rightarrow\infty}\mathbf{T}^{(k)}=c_{0}    
    \begin{pmatrix}
    1 \\
    \vdots \\
    1 
    \end{pmatrix}.
\label{eq:Tklimit-finished}
\end{equation}
This result is fundamental and can be interpreted as follows: the initial vector of the offsets $\mathbf{T}^{(0)}$, composed of unknown values, is iteratively transformed by the Markov chain, ``smoothing'' it out until the residual time offset is the same for all channels of the detector. This means that after a sufficient number of iterations $\mathcal{K}$, the time-corrected readout channels are synchronised, and the time calibration is complete. 

Equation~\ref{eq:Tklimit-finished} reflects another important feature of the problem. Since we always work with time differences between pairs of hits, this method has no information about the absolute time offset of each channel, but only the relative differences between channels. As a consequence, any solution that features a common time shift added on top of each channel's offset is an equally good time synchronisation. This \textit{common time offset} is a free parameter in our treatment, it does not affect the time calibration and can be arbitrarily fixed. For example, it can be fixed by setting the offset of a certain channel to an arbitrary value. We choose to not regard any channel as special, and fix the parameter by imposing that the average over all detector channels of the time corrections applied has to be zero at each iteration:
\begin{equation}
    \sum_{i=1}^{N_{ch}}\Delta^{(k)}_{i}=0\hspace{5mm}\forall k.
    \label{eq:fix_reference_time}
\end{equation}
This is equivalent to imposing that the computed time offsets average to 0 among all detector channels.

It is worth noticing that the common time offset is relevant only if the detector is used synchronously with other detectors, in which case an inter-detector time calibration is required and it is independent of the internal calibration of the readout channels of each detector. By applying this requirement, it can be seen that the only possible solution for equation~\ref{eq:Tklimit-finished} is when $c_0$ = 0, hence the limit in equation~\ref{eq:Tklimit-finished} converges to 0.

It is finally straightforward to show that the time offset that we estimate for each channel, defined in equation~\ref{eq:defOffsetEstimation_vector}, corresponds to $\mathbf{T}^{(0)}$.

According to equation~\ref{eq:correction_on_deltaT}, the correction applied at iteration $k$ can be written as 
\begin{equation}
    -\alpha\mathbf{\Delta}^{(k)}=\mathbf{T}^{(k+1)}-\mathbf{T}^{(k)}.
\end{equation}
Therefore, the estimated time offset vector after $\mathcal{K}$ iterations is:
\begin{equation}
    \mathbf{W}=\sum_{k=0}^\mathcal{K}\alpha\mathbf{\Delta}^{(k)} = -\left(\sum_{k=0}^\mathcal{K}\mathbf{T}^{(k+1)} -\sum_{k=0}^\mathcal{K}\mathbf{T}^{(k)} \right)= -(\mathbf{T}^{(\mathcal{K}+1)}-\mathbf{T}^{(0)}),
\end{equation}
which tends to the desired $\mathbf{T}^{(0)}$ in the limit of a large number of iterations, modulo the common time offset. So, the sum of all the corrections applied at each iteration compensates for the time offset of each channel.

\subsection{Rate of convergence}
One advantage of this method is that the dynamics of the algorithm is fully described by the matrix $M$ (i.e.\ the crossing matrix $\mathbb{X}$ along with the damping factor $\alpha$). Therefore, after having built the matching hits dataset, it is possible to infer properties about the algorithm's convergence by spectral analysis of $M$. 
In a Markov chain, the rate of convergence is measured by the spectral gap $\gamma=1-|\lambda_1|$, where $\lambda_1$ is the second largest eigenvalue of the transition matrix. The larger the spectral gap is, the faster the chain mixes. In our case, this means that the closer the second eigenvalue of $M$ is to 1, the quicker the vector of time offsets approaches the desired vector $\mathbf{T^0}$. Considering our transition matrix $M$, its second eigenvalue can be computed to first order approximation as 
$$ 
\lambda_1(M) = (1-\alpha) + \alpha\cdot\lambda_1(\mathbb{X}).
$$
Hence the spectral gap is 
$$ 
\gamma = \alpha[1-\lambda_1(\mathbb{X})],
$$
where $\lambda_1(\mathbb{X})$ is the second largest eigenvalue of the crossing matrix.
We notice that the spectral gap is proportional to $\alpha$, showing that the damping factor determines the convergence rate of the time calibration. Moreover, in this approximation, the second eigenvalue depends only on the crossing matrix $\mathbb{X}$, showing that the geometry of the detector, with regard to the way in which matching hit pairs are defined, drives the choice of $\alpha$ to obtain the desired rate of convergence. In general, it is possible to numerically compute the spectral gap $\gamma$ of a specific matching hit pairs dataset, to exactly infer the convergence rate in any specific case.
\label{sec:math}

\subsection{Algorithm implementation}
In a real case scenario, there is no need to explicitly compute the eigenvalues of the matrix, which might be computationally expensive if the number of readout channels is large. Instead, the iterative feature of the time offset compensation method makes it relatively easy to implement. The iterative procedure, sketched in pseudo code~\ref{algo}, is outlined here:  
\begin{enumerate}  
    \item Build the dataset of matching hits;  
    \item Loop through the dataset and compute $\Delta$ for each hit pair as in equation~\ref{eq:delta_n_definition}; 
    \item For each channel $ch_i$, compute $\Delta_i$ as the average of all $\Delta$ values in the dataset that include a hit read by channel $ch_i$;  
    \item Construct a vector $\mathbf{\Delta}$ with $N_{ch}$ elements. The elements of $\mathbf{\Delta}$ correspond to those defined in equation~\ref{eq:correction} for the first iteration; 
    \item Define the correction vector for this iteration as $\alpha\mathbf{\Delta}$, where $\alpha$ is the damping factor defined above;
    \item Repeat the procedure for a number of iterations, using the corrected time for each matching hit pair to compute $\Delta$, until the convergence criteria are met.  
\end{enumerate}  

Since the corrections applied at each iteration tend to 0, the number of necessary iterations can be evaluated based on the magnitude of $\alpha\mathbf{\Delta}$. The algorithm stops when 
\begin{equation}
     |\mathbf\Delta^k|< T_{\text{converge}}
\end{equation}

for a certain iteration $k$, where $T_{\text{converge}}$ is much smaller than the expected time resolution of the detector, ensuring that additional iterations will not significantly alter the result. In other words, the contribution of additional iterations would be negligible compared to the detector's time resolution.

\begin{algorithm}
\caption{Iterative time offset compensation}\label{algo}
{Dataset: [$t_1$,$t_2$,$s_{12}$,$ch_1$,$ch_2$]$_n$ for $n=1 $ to $N_{\text{matching hits}}$ }

\begin{algorithmic}[1]

\While{ $|\text{offset}|>T_{converge}$ }

\For{$n=1 $ to $N_{\text{matching hits}}$} \Comment{Loop through the matching hit pairs}
    \State $t_1 = t_{1n} - \text{offset}[ch_1]$
    \State $t_2 = t_{2n} - \text{offset}[ch_2]$
    \State $\Delta$ = $\alpha[(t_1 - t_2) - s_{12}]$ \Comment{$s_{12}$ is the expected time difference}
    \State $\text{offset}[ch_1]$ += $\Delta$
    \State $\text{offset}[ch_2]$ -= $\Delta$
    \State $N_{entries}[ch_1]$ += 1
    \State $N_{entries}[ch_2]$ += 1
\EndFor

\For{$ch=0$ to $N_{ch}$}    \Comment{Divide by number of entries in each channel}
    \State $\text{offset}[ch]$ /= $N_{entries}[ch]$
\EndFor

\For{$ch=0$ to $N_{ch}$}  \Comment{Shift all offsets to impose average 0 over channels}
 \State $\text{offset}[ch]$ -= $\langle \text{offset} \rangle$
\EndFor
\EndWhile

\end{algorithmic}
\end{algorithm}
\label{sec:algorithm}

\section{Application to the SuperFGD}
In the SuperFGD the scintillation light is conveyed through the optical fibres and read by around 56,000 SiPMs, arranged in PCBs of 64 sensors each. Several coaxial cables of different lengths convey the analogue signal to the front-end electronics system, where it is amplified and digitised by 222 Front-End Boards (FEBs). Each FEB reads 256 channels using eight separate readout chips; FEBs are organised in 16 separate crates. The signal time stamp is sampled with 400~MHz frequency for each channel of an FEB. Synchronisation signals are delivered to all the FEBs from a Main Clock Board, generating the global clock, and connected to all crates. It is evident that small time mis-synchronisation can happen at several stages along the readout chain, from imperfections in PCB trace-matching  and cable length measurements, to delays in the synchronisation of specific subsets of the electronics system, such as entire crates, FEBs, or readout chips. The goal of the time calibration of the SuperFGD is to find a fixed time correction offset to be applied to each of the 56,000 channels that accounts for all of these effects, effectively synchronising the system:

\begin{equation}
    T^{(0)}_i= \delta T_{\text{cable}} + \delta T_{\text{traces}} + \delta T_{\text{sync}}
\end{equation}
with $i=1,...,N_{ch}$.\\

In this section the iterative time compensation method is applied to the SuperFGD case. First, a simulation specific to the SuperFGD geometry is set up, to numerically validate the method with a smaller number of channels. 
Then, a set of SuperFGD matching hits obtained from a selection of data triggered on cosmic muons is used to perform the calibration of the detector channels, resulting in improved detector timing resolution.
\subsection{Geometrical model}
As a proof of concept, we made a simplified model of the SuperFGD readout geometry, that consists of $N_x\times N_y\times N_z$ cubes, crossed by optical fibres (``\textit{Toy-SFGD''}). As in the SuperFGD, each optical fibre crosses multiple cubes, depending on the fibre direction. We simulate a scintillating light signal simply as a hit generated in a random cube, and propagate the light assuming constant speed in the fibre. For each light signal, three hits are generated in the three corresponding fibres, out of which three matching hit pairs are built. With this simplified model, the measured time difference in each matching hit pair is equal to the expected time difference, estimated as $(d_1-d_2)/(v_\text{{fibre}})$.

To demonstrate that the method outlined in this manuscript finds the desired set of time offsets, we define for each channel a fixed time offset that is added to each time measurement, then we run the algorithm outlined in section~\ref{sec:algorithm}, and we compare the resulting vector of offsets with the input one. We define the input time offset vector by throwing random values from a probability distribution in a range of a few ns, and by imposing that the time offsets average out to 0 over all channels.

\begin{figure}
    \centering
    \includegraphics[width=1\linewidth]{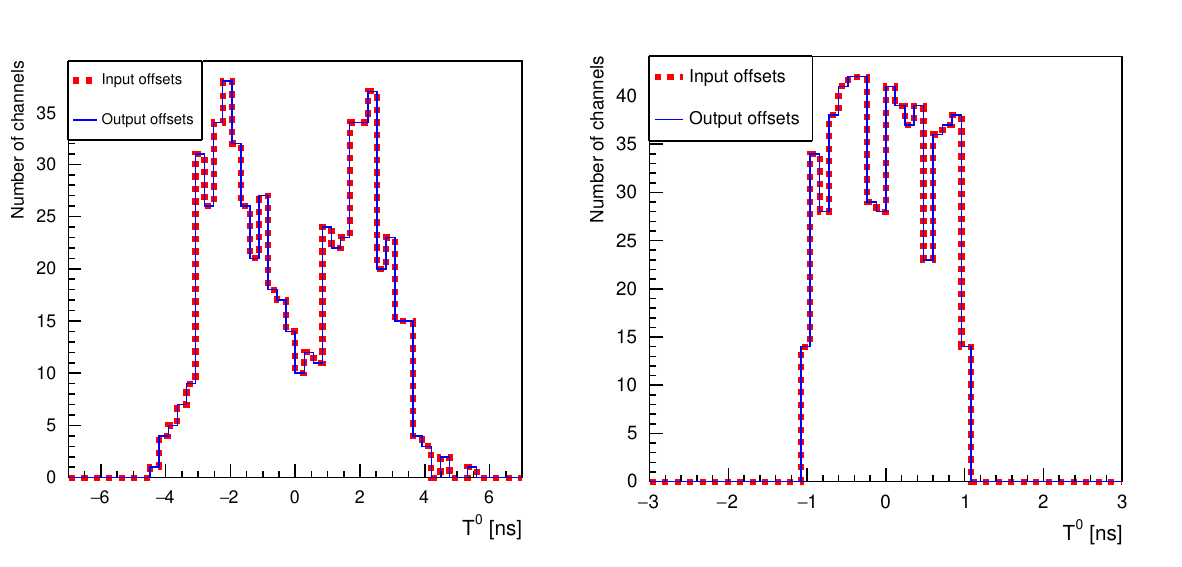}
    \caption{Distributions of input time offsets and the time offsets found by the time calibration algorithm (``output'' offsets), for two different configurations. Left: for each channel, the time offset is randomly drawn from two gaussian distributions at a distance of 2~ns, with $\sigma=1$~ns. Right: for each channel, the time offset is drawn from a uniform distribution in the interval [-1~ns, 1~ns].}
    \label{fig:input_output_offsets_numerical}
\end{figure}

To simplify the calculations, the simulated \textit{Toy-SFGD} consists of $20\times5\times20$ cubes, for a total of 600 readout channels. We study the evolution of the computed time offsets over 40 iterations. Figure~\ref{fig:input_output_offsets_numerical} shows that the outlined algorithm exactly finds the time offsets set in the simulation for all 600 channels for different offset configurations.

\begin{figure}
    \centering
    \includegraphics[width=0.5\linewidth]{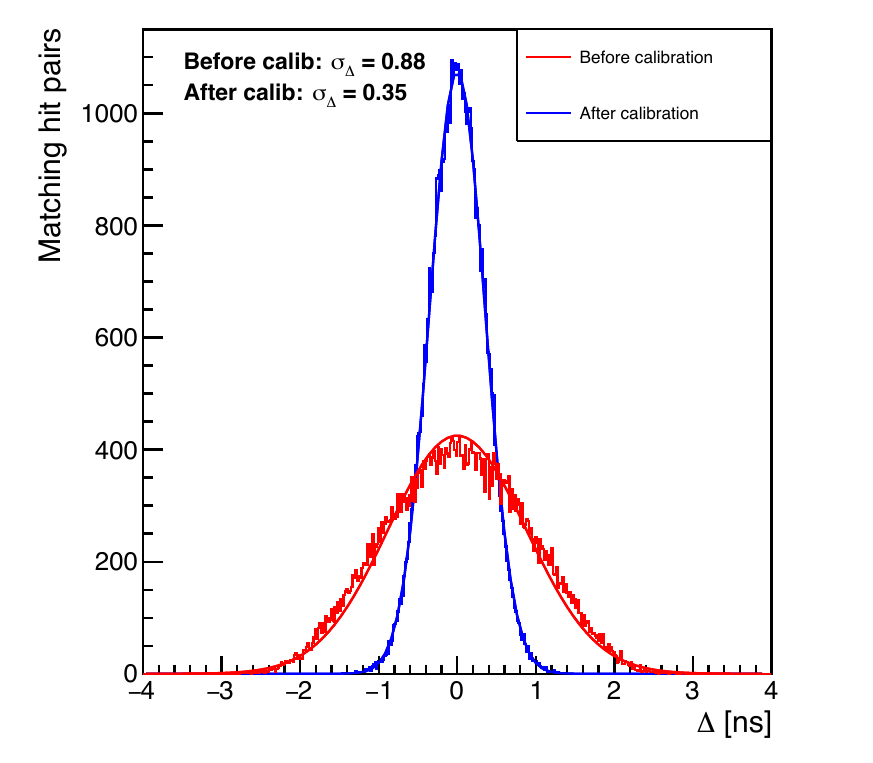}
    \caption{Distribution of $\Delta=(t_1-t_2)-(s_1-s_2)$ over the full set of matching hit pairs before and after the time offset calibration. The standard deviation of this distribution is linked to the intrinsic time resolution by: $\sigma_{\Delta}=\sqrt{2}\sigma_{t}$.}
    \label{fig:input_output_offsets_numerical_smeared}
\end{figure}

\begin{figure}
    \centering
    \includegraphics[width=0.95\linewidth]{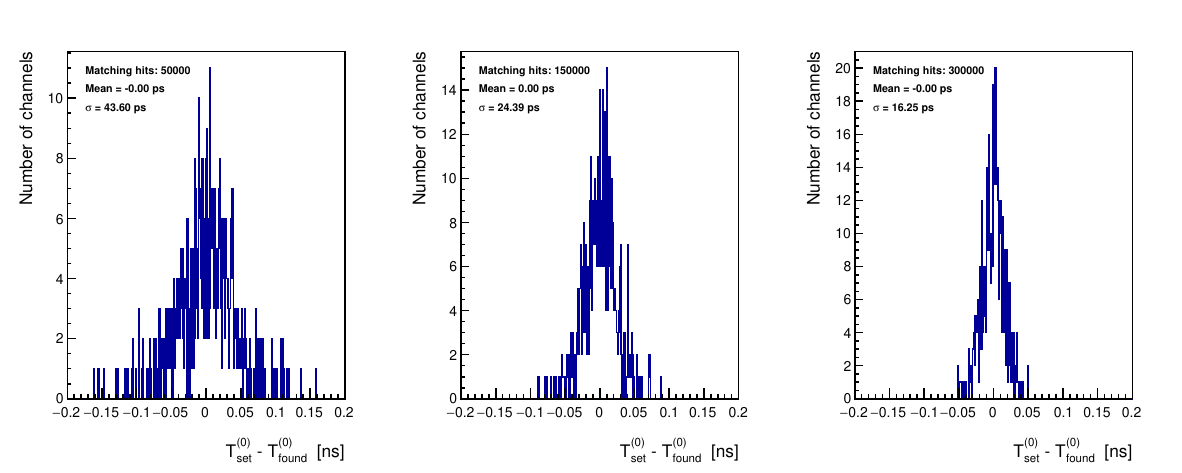}
    \caption{Distribution of the difference between the input time offset of a channel and the one estimated with the outlined method, for 3 matching hit pairs samples of different sizes (50,000, 150,000 and 300,000).}
    \label{fig:in_out_difference_250ps}
\end{figure}

To make the model more realistic, we can introduce a time smearing of 250~ps on each time measurement to reproduce the intrinsic time resolution of the detector, due to effects such as discrete time stamping, scintillation and photon travel time. Evidently, this will generate a smearing on the distribution of the difference between input and output offsets, and it will represent a lower limit of the time resolution after the calibration. In the ideal case where the uncertainties on the speed of light in the optical fibres and on the distances $d_1,d_2$ are negligible, of which this simulation is representative, the single channel time resolution can be estimated as $\sigma_t=\sigma_{\Delta}/\sqrt{2}$. Figure~\ref{fig:input_output_offsets_numerical_smeared} shows that the time resolution of the single channel, estimated as $\sigma_{\Delta}/\sqrt{2}$, reaches its minimum value after the time calibration has been performed: the detrimental effect of the channel time offsets is completely suppressed. In the case of finite time resolution, the precision of the time offset estimation of each channel depends on the number of matching hits in each channel, as shown in Figure~\ref{fig:in_out_difference_250ps}.


\begin{figure}
    \centering
    \includegraphics[width=0.5\linewidth]{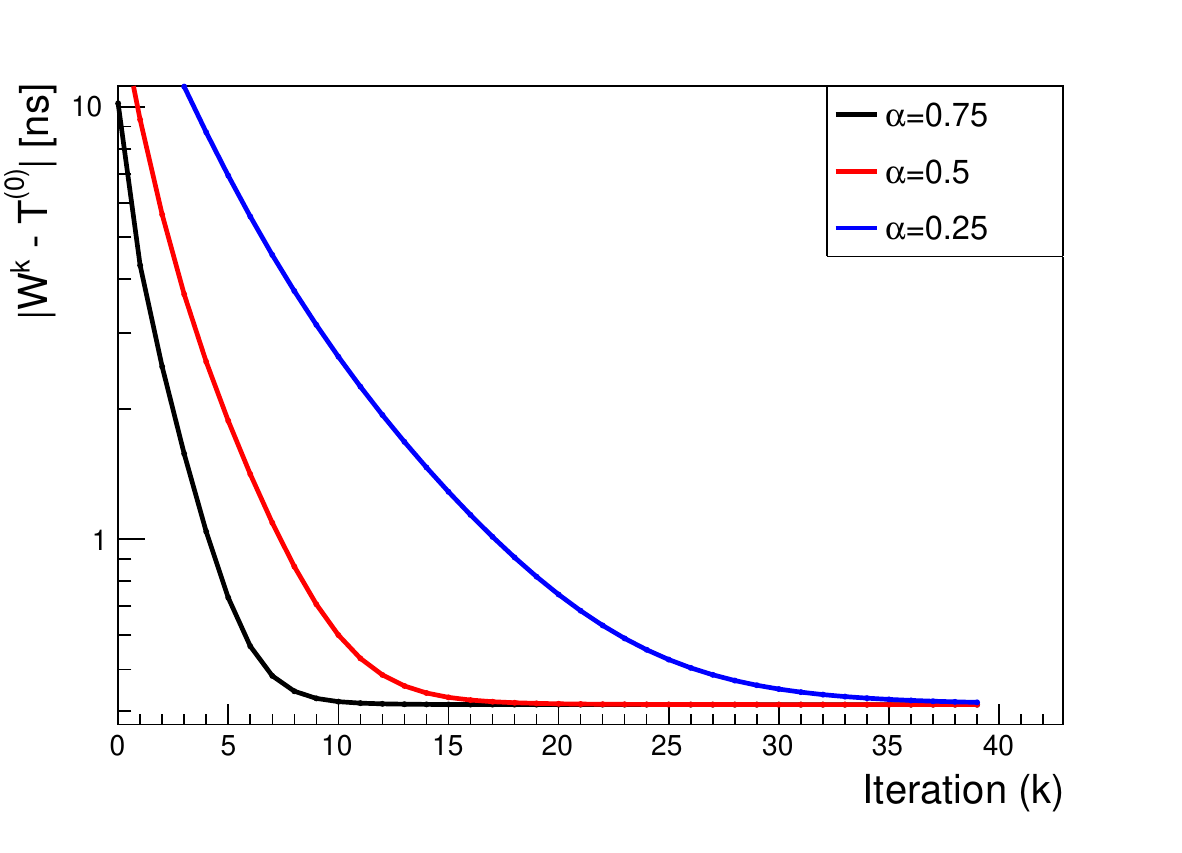}
    \caption{Distance between computed time offsets and input time offsets as a function of the iterations. $\mathbf{W^k}$ represents the vector of computed time offsets after $k$ iterations, and $\mathbf{T^0}$ is the input time offsets vector. Notice that $|\mathbf{W^k}-\mathbf{T^0}|$ does not tend to zero due to the finite intrinsic time resolution $\sigma_t=250~\text{ps}$. }
    \label{fig:convergence_speed_numerical}
\end{figure}

Finally, by looking at how fast the computed time offsets approach the input time offsets, this simple model allows us to understand the rate of convergence. Figure~\ref{fig:convergence_speed_numerical} shows the different convergence speed as the damping factor $\alpha$ varies.
  \label{sec:toySFG}
\subsection{Results with calibration data}
The iterative time offset compensation method outlined in the previous section was used to perform the time calibration of the SuperFGD detector. Detector commissioning data was used to obtain equal numbers of matching hit pairs from samples of cosmic ray muons and neutrino beam interactions. Approximately 3 hours of cosmic muon data taking produces 5 million matching hits. Another 5 million matching hits can be obtained from 36 hours of continuous T2K neutrino beam data. The speed of light propagation in the optical fibres is assumed to be 167~mm/ns, obtained from internal measurements.

\begin{figure}
    \centering
    \includegraphics[width=0.73\linewidth]{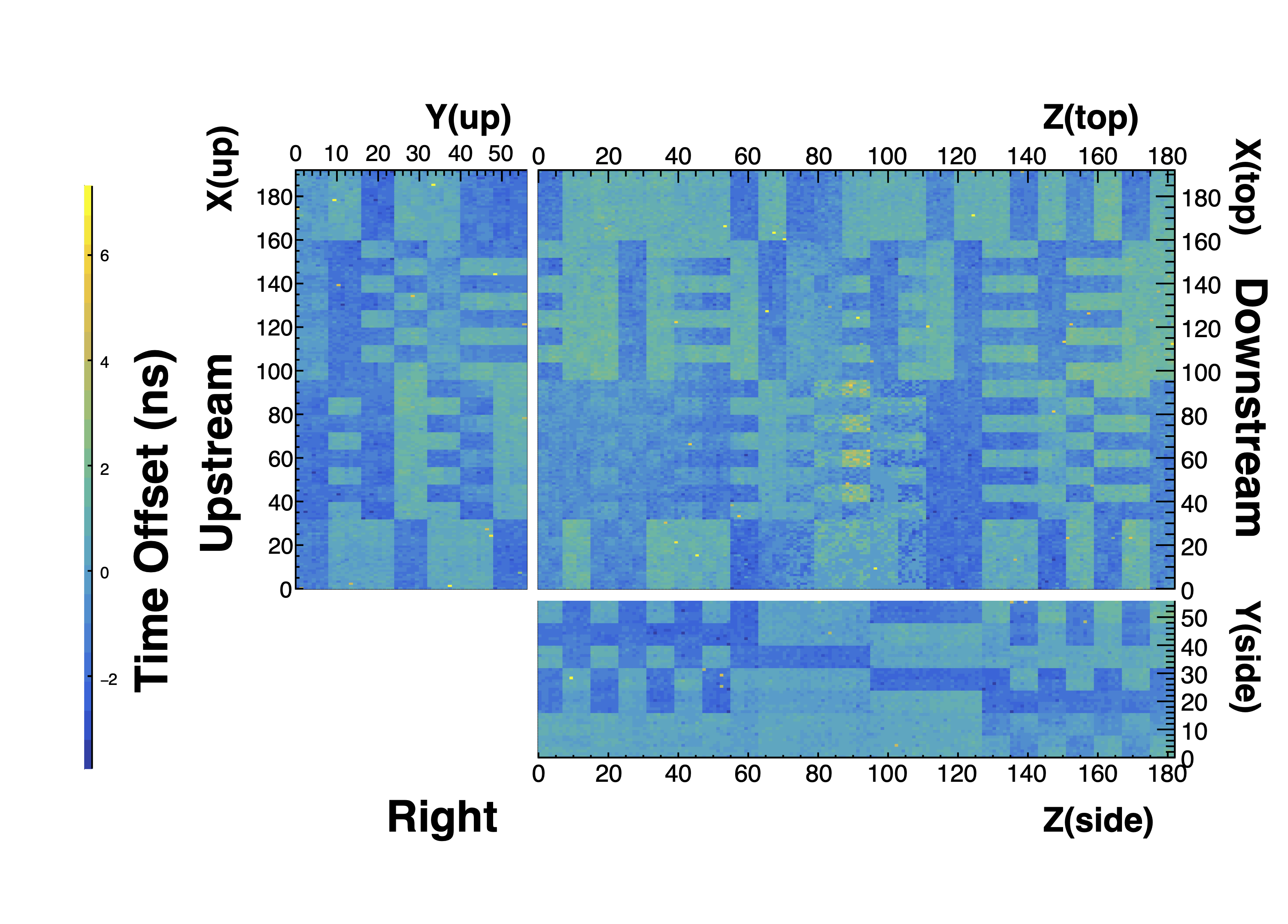}
    \caption{Channel map of the estimated time offsets for the SuperFGD. The separation between different electronics units (FEBs) affected by different time delays is evident. }
    \label{fig:offsetMap}
\end{figure}

The above-described matching hit pair sample is used to estimate the time offset for all the detector channels in the SuperFGD. The resulting per-channel offset is shown in Figure~\ref{fig:offsetMap}. The obtained offsets are used to calibrate an independent validation dataset of matched hit pairs, obtained from 24 hours of continuous neutrino beam.

\begin{figure}
    \centering
    \includegraphics[width=0.6\linewidth]{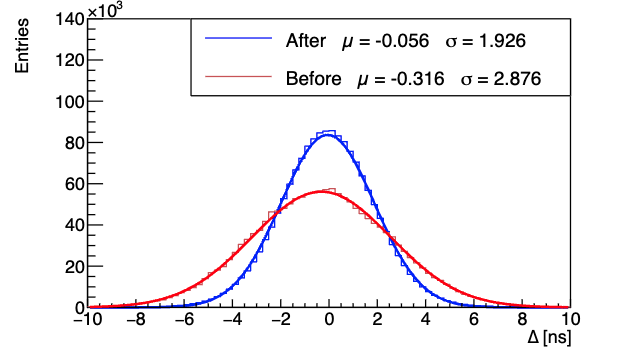}
    \caption{Distribution of $\Delta=(t_1-t_2)-\dfrac{d_1-d_2}{v_\text{{fibre}}}$ among the matching hit pairs of the validation dataset before and after applying the time offset compensation. Hits with charge larger than 20 p.e.\ are selected to ensure rejection of noise hits.}
    \label{fig:before_after_calibration}
\end{figure}

We can estimate the average time resolution of the single detector channel using the distribution of $\Delta=(t_1-t_2)-\dfrac{d_1-d_2}{v_\text{{fibre}}}$ among the validation dataset. If the expected time difference between pairs of hits is known with enough precision, we can assume that the spread on $(d_1-d_2)/v_\text{{fibre}}$ is negligible compared to that of $t_1-t_2$. In this case, the time resolution of single readout channels can be estimated as $\sigma_t\simeq\sigma_\Delta/\sqrt2$. Figure~\ref{fig:before_after_calibration} shows the effect of the time offset compensation on the time resolution as measured in the validation dataset. Considering that the typical light yield of a minimum ionising particle (MIP) in the SuperFGD is around 30-40~p.e., the validation dataset includes only hits yielding an SiPM charge larger than 20 p.e., in order to reject noise hits.
The iterative time offset compensation reduces the estimated time resolution of the single channel from $1.81$~ns to $1.36$~ns.
\label{sec:data}
\newpage
\section{Application to the Time Of Flight detector}
The ToF detector presents another interesting case. The time correlation between pairs of hit times is caused by the time of flight of cosmic muons between pairs of scintillating bars in different planes. Therefore, the Markov matrix that defines the dynamics of the synchronisation procedure (computed with the general definition in equation~\ref{eq:definition_M_matrix}), looks very different with respect to the Super-FGD case, not changing the validity of the proof, but showing the effectiveness of the treatment in a system with different dynamics. The time calibration discussed in this work relies on the capability of the ToF detector to estimate the position along the bar where the particle has generated scintillating light. Therefore, we note two constraints important to the application of the calibration procedure to the ToF detector:
firstly, only the reconstructed interaction time is affected by the iterative time correction, while the reconstructed position along the bar is fixed throughout the calibration procedure.  
Secondly, only the ToF signals with a coincidence of hits on both ends of the bar are used. This is necessary because only with two valid time readouts at both ends of a scintillating bar can the ToF estimate the hit position.

For these reasons, during the ToF calibration it is assumed that relative time shifts between the two readout channels on each end of the bar can be neglected, and the time offset $T^{(0)}_i$ (assigned to the $i$-th bar) computed with this calibration procedure identifies an average time shift of the readout devices of each bar. Moreover, $T^{(0)}_i$ not only contains the time delays due to the electronics readout mis-synchronisation, but will also absorb any geometric variations with respect to the reference geometry:
\begin{equation}
    T^{(0)}_i = \dfrac{T^{(0)}_{Li}+T^{(0)}_{Ri}}{2} = \delta T_{cables} + \delta T_{geometry},
\end{equation}
where $i=1,...,N_{bars}$, while $T^{(0)}_{Li}$ and $T^{(0)}_{Ri}$ are the specific time offsets of the readout channels on the two ends of each bar, an additional calibration procedure is required to calibrate their difference.

In this section, the time compensation method is validated on a geometrical model simulating the ToF detector. Then, ToF calibration is performed by means of the iterative time compensation method using a sample of more than 180,000 cosmic ray muons, demonstrating an improvement of the detector time resolution. 
\subsection{ Geometrical model }
A simplified model of the ToF detector with five bars per panel and with a bar length of 2~m was used to numerically validate the time calibration method. The ToF panels are created as six sides of a cubic volume  with the center of the cube being the origin of the reference system. 
In this simulation, the ToF dual-ended bar readout is simplified with a single readout channel per bar, providing hit timing information and position along the bar, in accordance with the assumptions made for the calibration process. 
A pair of simulated matching hits represents a particle crossing two bars of different panels, with the initial hit in the first bar always at time $t_0 = 0 $ and the second hit at a time $t_1 = d/c$, where $d$ is the distance between the two hit positions and $c$ is the speed of light. The hits are randomly generated at different longitudinal positions along the bars.

\begin{figure} [H]
    \centering
    \includegraphics[width=0.6\linewidth]{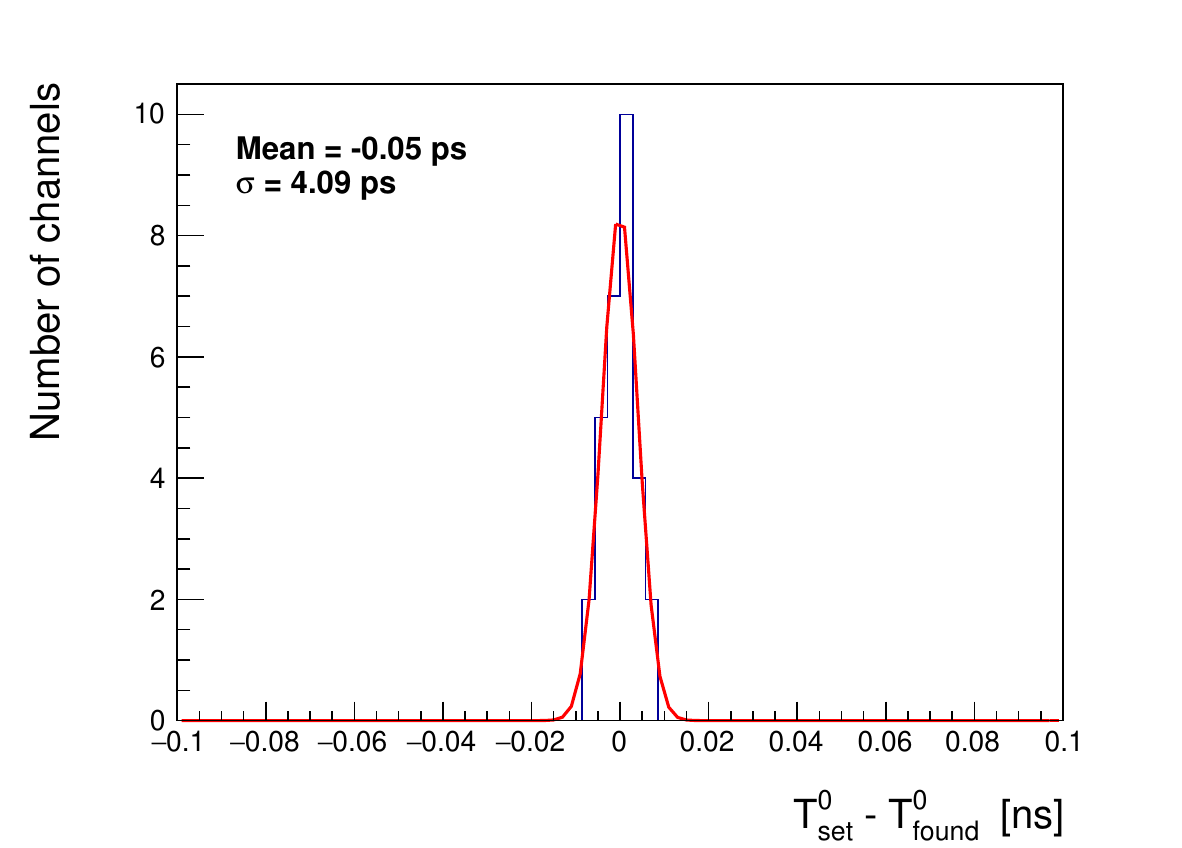}
    \caption{Distribution of the difference between the input time offsets set in the simulation and the time offsets estimated with the calibration method.}
    \label{fig:tofOffset_before_after}
\end{figure}

Similar to the SuperFGD case described in section~\ref{sec:toySFG}, a time offset is defined for each channel and added to the hit times during hit generation. The matching hit pairs sample is processed through the calibration algorithm, and the estimated offsets are compared to the input ones. Random offsets are generated between -1~ns and 1~ns, with the imposition of an average offset of 0 over all channels. A smearing of 130~ps, which is the benchmark for timing resolution of the ToF detector~\cite{Korzenev:2021mny}, was added to make the simulation realistic, and will act as the lower limit for the calibration procedure.

\begin{figure}[!h]
    \centering
    \includegraphics[width=0.6\linewidth]{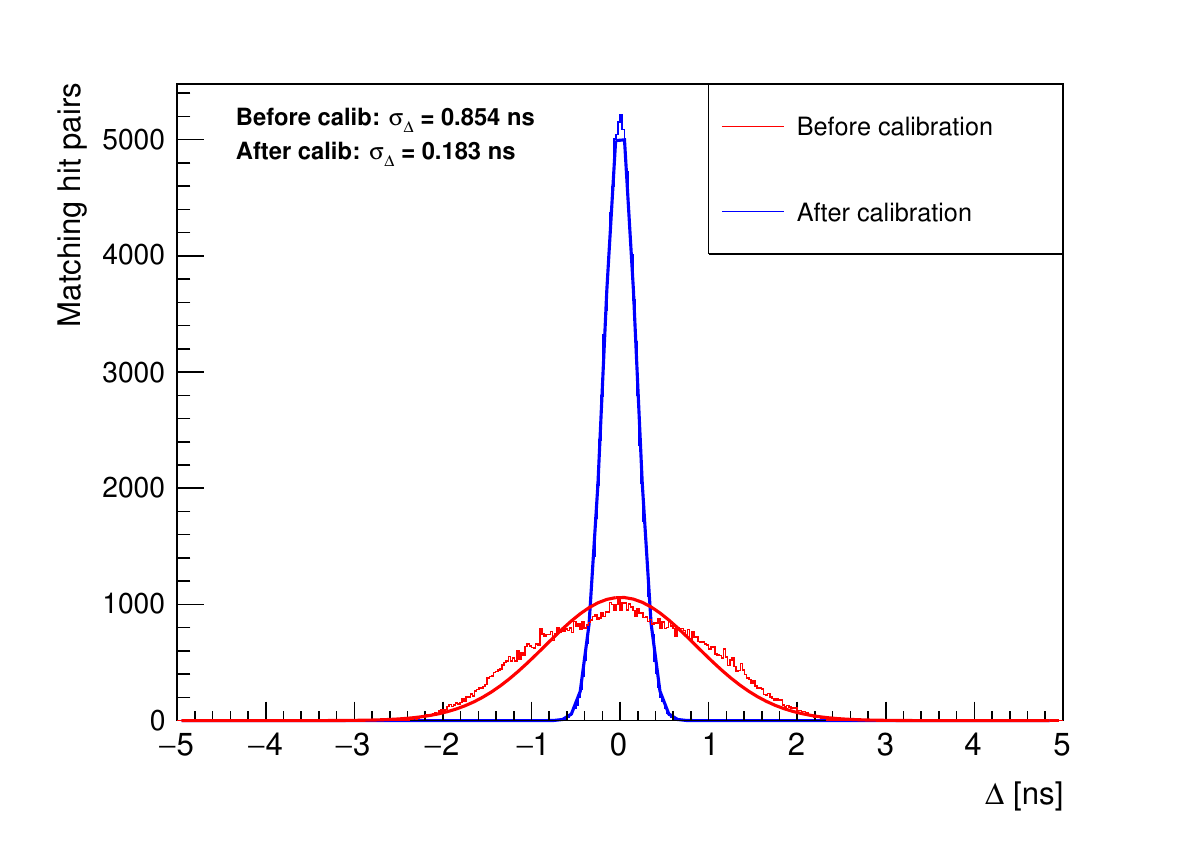}
    \caption{Time resolution estimated by $\Delta$ before and after the calibration with input intrinsic resolution = 130~ps. Note that the time resolution $\sigma_t$ is estimated as the standard deviation of the $\Delta$ distribution divided by $\sqrt{2}$.  }
    \label{fig:time_res_before/after_tof}
\end{figure}

The results after 100 iterations are shown in Figure~\ref{fig:tofOffset_before_after} with the algorithm finding the offsets very close to those set as inputs to the simulation.
The time resolution per channel is again estimated as $\sigma_t = \sigma_\Delta / \sqrt{2}$ which reaches the set value after the process is complete, as shown in Figure~\ref{fig:time_res_before/after_tof}. 
\label{sec:toyToF}
\subsection{Results with calibration data}  
\begin{figure} [!h]
    \centering
    \includegraphics[width=1\linewidth]{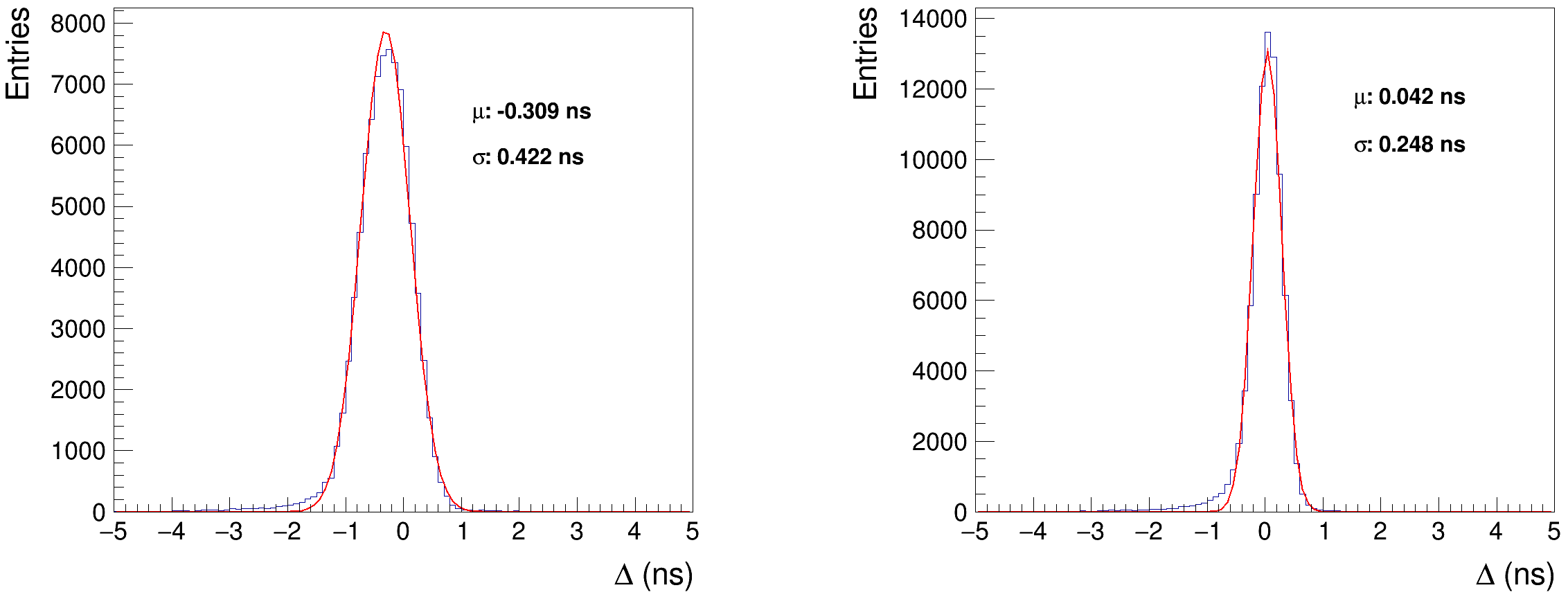}
    \caption{Improvement in time resolution for particles crossing top and bottom panels. The left plot shows the time resolution before corrections and the right plot shows after.}
    \label{fig:Time_resolution_top_bottom}
\end{figure}

\begin{figure} [!h]
    \centering
    \includegraphics[width=1\linewidth]{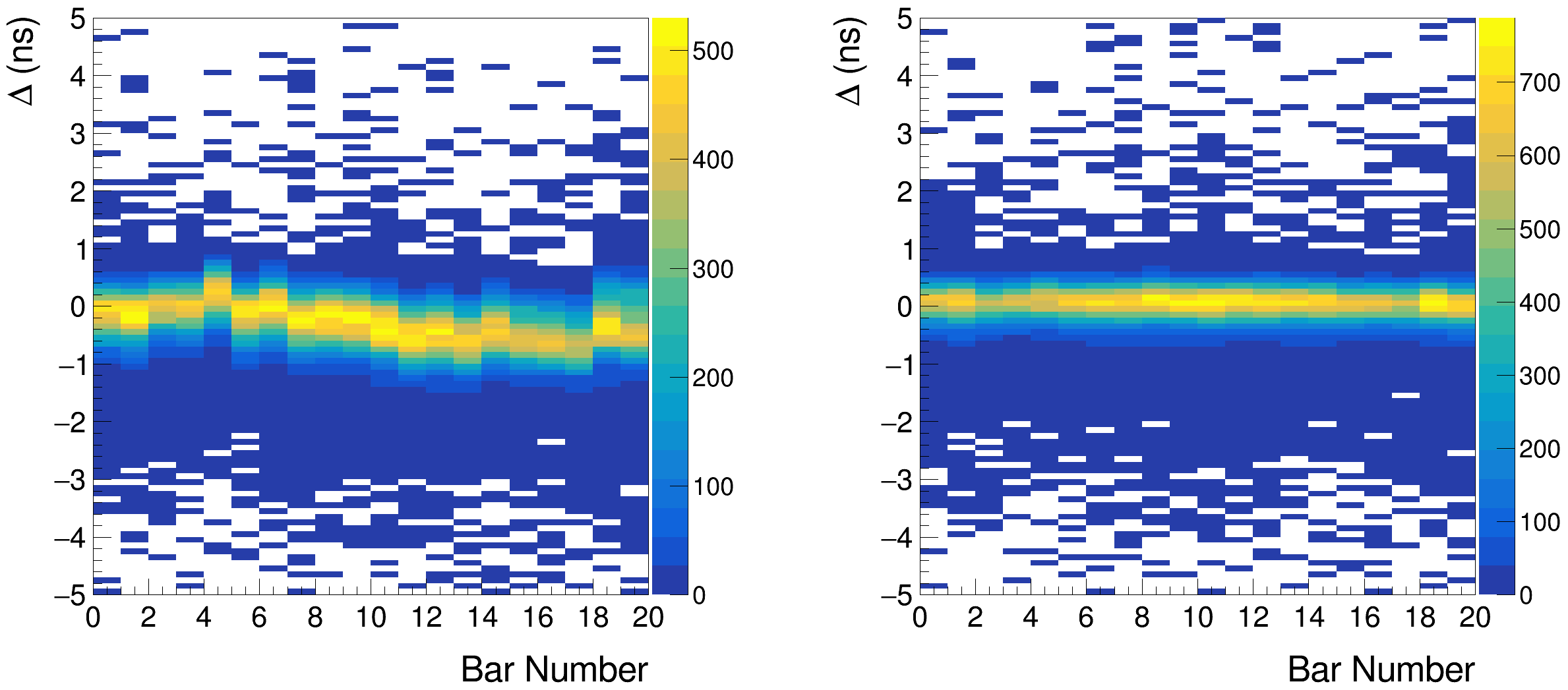}
    \caption{Delta times ($t_1-t_2-d/c$) computed for a given top bar in combination with all the bottom bars it can cross. The improvement can be seen on the right with variations reduced.}
    \label{fig:Tof_bar-bar_variation}
\end{figure}

Cosmic muon data was used to measure the offsets for each channel of the detector. In a typical cosmic muon run, due to the angular distribution of cosmic rays, most tracks produce one hit in the top panel and one hit in one of the other five panels. As a consequence, top-crossing matching hit pairs dominate the dataset. The matching hit pairs dataset can be further selected to increase balance between hits in different panels.
As in the geometrical model, the time resolution in data is estimated by dividing the standard deviation of a gaussian fit of the $\Delta$ distribution by $\sqrt{2}$. We observe an improvement in the time resolution from 298.2 ps to 175.3 ps in the top-bottom tracks, as shown in Figure~\ref{fig:Time_resolution_top_bottom}. Moreover, the mean of $\Delta=t_1-t_2-\frac{d}{c}$ between bars belonging to the top and bottom panels is shifted towards zero thanks to the minimisation procedure, as it can be seen in the bar-to-bar variations in Figure~\ref{fig:Tof_bar-bar_variation}. 
It is important to note that the precision of this method depends on the position resolution, limited to approximately $3$~cm in the case of the ToF detector~\cite{Korzenev:2021mny}. The calibration performance using the same method, and consequently the ToF time resolution, can be further improved if an external detector with better position resolution is used to infer the position of the interaction point along the bars.
\section{Conclusions}
The current and next generation of particle physics experiments require precise time synchronisation across large-scale detector systems with numerous readout channels. In this work, we have introduced a novel time calibration method based on Markov Chains, which effectively determines inter-channel time offsets due to mis-synchronisation, using only the intrinsic time correlations between pairs of detected hits. This approach eliminates the reliance on external reference time signals, making it particularly advantageous for detector setups where the track reconstruction is not precise enough or computationally very expensive. Moreover, the outlined method provides a handle on the convergence speed through the damping parameter $\alpha$, making it suitable for setups with a very different number of readout channels, whilst maintaining a reasonable convergence time.

Through numerical simulations, we have demonstrated the robustness and accuracy of our method in recovering time offsets across detector channels. Furthermore, we have validated its effectiveness in existing detector concepts by applying it to the SuperFGD and ToF detectors at the T2K near detector, showing that it is possible to improve the time resolution in profoundly different detector setups with the same method. The results confirm that our approach can successfully mitigate clock-phase effects and hardware trace-length mismatches, thereby enhancing overall detector timing performance. It must also be noted that the calibration procedure is perfectly scalable to systems with smaller single-channel time resolution, not relying on external time references that might limit the time calibration capabilities.

More generally, this method can be used to enhance the time performance of any time-synchronised system, correcting for different time-shifting effects. As an example, a minimal extension of this method can be used to correct for the time-walk effect in scintillator detectors, by allowing the time offset $T^{(0)}$ of each channel to vary as a function of the recorded charge. Moreover, it can be used to correct for light propagation effects, such as the effective speed of light along optical fibres, by including the estimated distance between the scintillating light emission and the light readout device as an additional degree of freedom. In these extensions of the method, the only difference resides in the size of the $T^{(0)}$ vector, or in a replacement of the constant terms in $T^{(0)}$ with a function of other relevant physical parameters.

By providing a scalable and detector-independent solution to inter-channel mis-synchronisation, this method is well suited for particle detectors employing scintillation-based ionisation detection. Its applicability extends beyond scintillation-based ionisation detection, offering a general framework for improving timing precision in detector systems with large numbers of readout channels. A similar Markov Chain technique can be applied to inter-channel calibration of energy response, provided there is a method to correlate energy deposition between detector channels, similarly to the way in which matching hit pairs samples are generated in this work.

The application of this method to the SuperFGD improves the average single fibre time resolution from $1.81$~ns down to $1.36$~ns for hits yielding more than 20~photoelectrons in the SiPM, enabling an estimated sub-ns time resolution for a cube hit ($\sigma_t[\text{cube}]\simeq\sigma_t[\text{fibre}]/\sqrt3$). A sub-ns time resolution allows the SuperFGD detector to measure the time of flight of neutrons produced in neutrino-nucleus interactions, enabling precise measurement of neutron kinetic energy~\cite{T2K:2019bbb}. 

The application of this method to the ToF detector improves the estimated time resolution of the single bar from 298.2~ps to 175.3~ps, crucial for the capability of the SuperFGD and ToF system to distinguish forward going and backward going tracks, and providing promising results for particle identification by time of flight. The same method can be applied to a dataset of matching hit pairs obtained with an external position reference to further improve the time resolution. 
\appendix
\section{Periodicity and reducibility of the Markov matrix}
\label{sec:irreducible}
In section~\ref{sec:math}, we used a result of a well-known theorem in algebra of Markov Matrix and stochastic processes, from Perron and Frobenius~\cite{Perron-Frobenius}. In this appendix we discuss the conditions under which this theorem applies in our specific case, in order to understand when these conditions are met, so that convergence of the algorithm can be demonstrated.

The Perron-Frobenius theorem states that the largest eigenvalue $\lambda_0$ in absolute value (called the Perron-Frobenius eigenvalue) of a real non-negative matrix $A$ of elements $a_{ij}$ is always real, and is such that:
\begin{equation}
    \min_{i}\sum_j a_{ij}<|\lambda_0|<\min_{i}\sum_j a_{ij}.
\end{equation}

In a row-stochastic matrix, this means that the Perron-Frobenius eigenvalue is 1. If this applies, and if, additionally, the matrix is aperiodic and irreducible, the matrix series 
$ \left(\sum_k A^k\right)$
converges. Since our matrix describes a defined physical problem, we can say something more about these properties.
\begin{enumerate}
\item {\bf Periodicity :}
Periodicity occurs when a Markov chain visits the same state every $n$ steps, with $n>1$. It is possible to show that a Markov chain described by a matrix where all the diagonal entries are strictly positive can not be periodic. By our definition of the matrix, $M$ is always aperiodic.
\item {\bf Reducibility : }
A matrix is said to be reducible if it can be transformed into an upper block-triangular matrix through simultaneous row and column permutations. This can also be visualised with the help of graph theory: a directed graph can be described by a matrix A, with $a_{ij}$ being different from 0 if there is a connection from node $i$ to node $j$ of the graph. A graph is said to be strongly connected if it is possible to move from any node to any other node following the connection lines. There is a bijective correspondence between irreducible matrices and strongly connected graphs. This gives an opportunity to visualise the channel intersections of a detector with a graph, to understand its reducibility properties.

In terms of matching hit pairs in a detector, being irreducible (or strongly connected) means that there is no subset of channels that forms matching hit pairs only among themselves, without intersecting the rest of the channels. We argue that, in most cases, the channel geometry of a detector with many channels features irreducibility. Moreover, in the case this condition does not hold for the whole detector, the detector would be divided into a (usually small) number of independent subsets of channels, that can be calibrated individually following the outlined method. Then, inter-calibration between the different subsets can be performed with a traditional method, or with the method outlined here if matching hit pairs between subsets can be defined.
\end{enumerate}

\section{Example of crossing matrices}
\begin{figure}[h!]
\centering
\includegraphics[width=0.45\columnwidth]{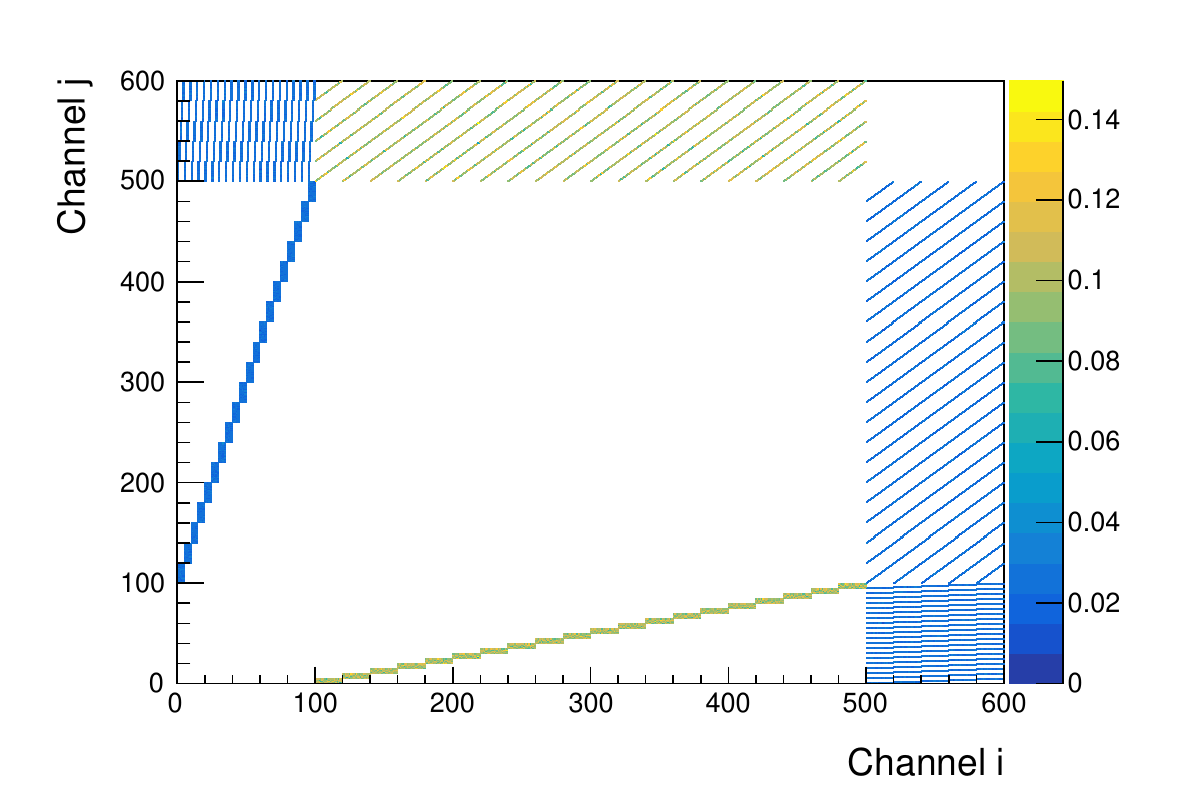}
\caption{
Crossing matrix $\mathbb{X}$ from a simulation of a SuperFGD-like detector geometry with 600 fibre readout channels: channels 0 to 99 read the XY projection, channels 100 to 499 read the XZ projection, and channels 500 to 599 read the YZ projection.
}
\label{fig:SFG_matrix}
\end{figure}

\begin{figure}[h!]
\centering
\includegraphics[width=0.45\columnwidth]{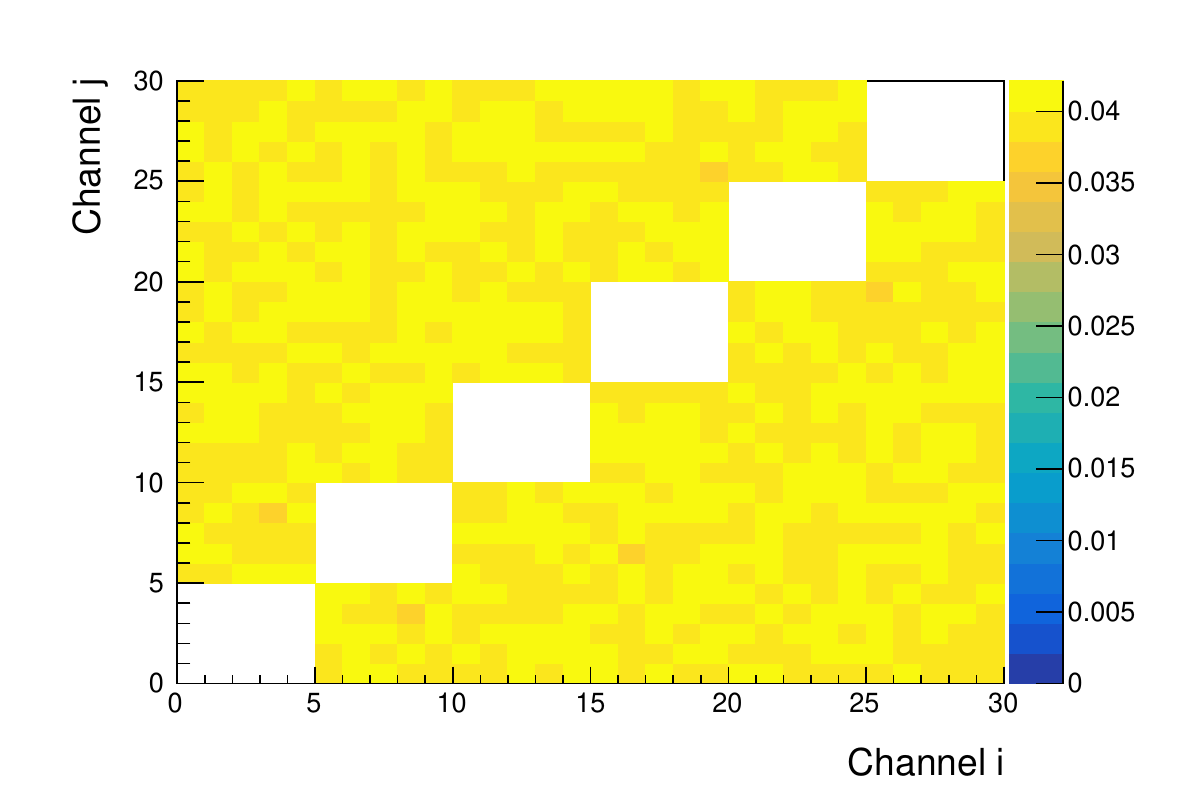}
\caption{Crossing matrix $\mathbb{X}$ from a simulation of a ToF-like detector geometry with 5 scintillating bars per plane.}
\label{fig:ToF_matrix}
\end{figure}

It was shown how the outlined iterative calibration method can be applied to various types of detectors with the only condition of defining a meaningful sample of \textit{matching hit pairs}. Any difference between detectors or between different definitions of matching hit pairs is encoded in the matrix that describes the Markov chain. In particular, we showed how the so-called crossing matrix $\mathbb{X}$, used to build the Markov matrix $M$ and naturally arising from the formalisation of the iterative calibration method, contains information about the matching hit pairs sample and detector geometry. 
It is therefore useful to look at the structure of the crossing matrix for different detector geometries, in the simple case of smaller-size ToF-like and SuperFGD-like detectors. Using the numerical models described in sections~\ref{sec:toySFG} and~\ref{sec:toyToF}, we constructed the crossing matrices $\mathbb{X}$, displayed in Figures~\ref{fig:SFG_matrix} and~\ref{fig:ToF_matrix}. Empty matrix entries represent pairs of channels for which a matching hit pair cannot exist, or is extremely unlikely, such as channels reading parallel fibres in the SuperFGD, or scintillating bars of the same plane in the ToF. In general, the magnitude of a matrix entry $x_{ij}$ is proportional to the likelihood of having a matching hit pair connecting channels $i$ and $j$, depending on the geometry of the detector with respect to the definition of matching hit pairs.

It is interesting to notice that the crossing matrix is not necessarily symmetric. In the case of the SuperFGD, although the location of non-zero entries is symmetric by definition of $\mathbb{X}$, the non-zero entries $x_{ij}$ and $x_{ji}$ must differ to satisfy row-stochasticity in the case of a non-cubic SuperFGD-like detector. Whether or not the crossing matrix is symmetric depends on the detector geometry.

\acknowledgments

This work was supported in the framework of the State project ``Science'' by the Ministry of Science and Higher Education of the Russian Federation under  contract No. 075-15-2024-541, and by the Swiss National Foundation Grants No. 200021\_204609 and 200021\_203433 and the DFG No. 517206441, Germany. We acknowledge the support of  JSPS KAKENHI Grant Numbers 16H06288 and 20H00149 and  the SNSF grant PCEFP2\_203261, Switzerland. We gratefully acknowledge the support of CNRS/IN2P3, France; DOE, USA; the STFC and UKRI, UK.  

In addition, participation of individual researchers has been supported by the European Union’s Horizon 2020 Research and  Innovation Programme under the grant number RISE-GA872549-SK2HK, and RISE-GA822070-JENNIFER2 and the Horizon Europe Marie Sklodowska-Curie Staff Exchange project JENNIFER3 Grant Agreement no.101183137, and by the Swiss-Vietnam  project no IZVSZ2.203433: Science Nation Foundation, Switzerland and NAFOSTED, Vietnam.

For the purposes of open access, the authors have applied a Creative Commons Attribution licence to any Author Accepted Manuscript version arising. Representations of the data relevant to the conclusions drawn here are provided within this paper.

\bibliographystyle{JHEP}
\bibliography{biblio.bib}

\end{document}